\documentclass{ametsocV6.1}
\nolinenumbers

\title{Nonlocal, Pattern-aware Response and Feedback Framework for Regional Climate Change}

\authors{Parvathi Kooloth,\aff{a}
\correspondingauthor{Parvathi Kooloth, parvathi.kooloth@pnnl.gov} 
Jian Lu,\aff{a} \correspondingauthor{Jian Lu, jian.lu@pnnl.gov} 
Yi Huang,\aff{b} 
Derek DeSantis,\aff{c} 
Yiling Huo,\aff{a} 
Fukai Liu, \aff{d} and
Hailong Wang \aff{a}
}

\affiliation{ \aff{a} Pacific Northwest National Laboratory, Richland, 99354 WA, 
 USA\\
\aff{b}McGill University, Montréal, H3A 0G4 QC, Canada\\
\aff{c}Los Alamos National Laboratory, Los Alamos, 87545 NM, USA\\
\aff{d} Ocean University of China, Qingdao, 266100, China}
\abstract{We devise a pattern-aware feedback framework for representing the forced climate response using a suite of Green’s function experiments with solar radiation perturbations. By considering the column energy balance, a comprehensive linear response function (CLRF) for important climate variables and feedback quantities such as moist static energy, sea surface temperature, albedo, cloud optical depth, and lapse rate is learned from Green’s function data. The learned CLRF delineates the effects of the energy diffusion in both the ocean and atmosphere and the pattern-aware feedbacks from the aforementioned radiatively active processes. The CLRF can then be decomposed into forcing-response mode pairs which are in turn used to construct a reduced-order model (ROM) describing the dominant dynamics of climate responses. These mode pairs capture nonlocal effects and teleconnections in the climate and thus, make the ROM apt for capturing regional features of climate change response. A key observation is that the CLRF captures the polar amplified response as the most excitable mode of the climate system and this mode is explainable in the data-learned pattern-aware feedback framework. The ROM can be used for predicting the response for a given forcing and for reconstructing the forcing from a given response; we demonstrate these capabilities for independent forcing pattern ($\approx 3.5$ W/m$^2$).} 

\begin{document}

\maketitle

%
%
%
\statement{Climate sensitivity and feedbacks have traditionally been examined as a zero-dimensional problem, neglecting the crucial patterns of response and forcing. This limitation has long hindered the acquisition of robust, patterned climate change information essential for informed decision-making. In this study, we strive to develop an innovative pattern-aware feedback framework by solving an inverse problem in a finite-dimensional space using a data-driven, optimization approach. The resultant reduced-order representation of the original system not only affords a feedback framework to quantify the patterns of the climate response but also reveals the most excitable---and thus most robust---modes of the climate system. The predictive power exhibited by the reduced-order model shows promise for optimizing climate forcing for geoengineering applications.  
}

%
%

%
\section{Introduction}

Climate sensitivity, the measure of how much the climate responds to a doubling of CO$_2$ concentration in the atmosphere, is a pivotal factor in shaping climate policy. It has garnered extensive attention and research within the climate science community, serving as a crucial benchmark for understanding the potential impacts of greenhouse gas emissions on global temperatures. The framework for quantifying climate sensitivity through feedback analysis draws inspiration from an electrical circuit (\cite{black1977inventing,roe2009feedbacks,goosse2018quantifying}), where the output of the system can influence its input, either positively reinforcing or attenuating the initial signal. For climate sensitivity (e.g., \cite{roe2009feedbacks}), with the global mean surface temperature as the response (or state) variable, the total response ($\Delta T$) to external radiative forcing ($\Delta R_f$) can be broken down as
\begin{equation}
    \Delta T = \frac{1}{\lambda_0} \frac{\Delta R_f}{\left(1 - \frac{1}{\lambda_0} \sum_i \lambda_i \right)} = \frac{1}{\lambda_0} \frac{\Delta R_f}{\left(1 -  \sum_i f_i \right)},
    \label{eqn:feedback}
\end{equation}
 where $\lambda_0$ is the reference-system climate sensitivity parameter, often chosen to be the Planck feedback, $\lambda_i$  is the feedback parameter with the corresponding feedback factor $f_i \equiv \lambda_i/\lambda_0$ indicating the fraction of the system’s output fed back into the input through each feedback process. Note here the sign convention for $\lambda_0$ is chosen to be positive, meaning a stabilizing/damping effect on the temperature anomaly under an external radiative forcing; the purpose of this choice will be clear later when the reference sensitivity is generalized to include the atmospheric energy transport effects. According to the sign convention of $\lambda_0$, positive $f_i$ implies positive feedback. For a zero-dimensional problem with known values of reference sensitivity ($\lambda_0$) and feedback parameters ($\lambda_i$), equation (1) allows for predicting the full response to any given forcing, or conversely, determining the forcing required to generate a specific response. When the formalism (\ref{eqn:feedback}) is applied to the global surface temperature response, which in turn is employed to diagnose global climate sensitivity and climate feedback in climate models, it is often used in combination with a radiative kernel (e.g., \cite{held2000water, soden2008quantifying, shell2008using, colman2009climate, zelinka2012climate, jonko2012climate, pendergrass2018surface, smith2020hadgem3}). The radiative kernel is derived from a Taylor series expansion of the radiative adjustments caused by feedbacks with respect to a specified reference system
 \begin{equation}
     \lambda_i = \frac{\partial R}{\partial \alpha_i} \bigg |_{\alpha_j, j\neq i} \frac{d \alpha_i}{dT},
     \label{eqn:f_mean}
 \end{equation}
where $\frac{\partial R}{\partial \alpha_i}$ is a kernel matrix often calculated as the radiative flux per unit change in the feedback variable at each model grid cell of the model, $d \alpha_i/dT$ is a vector when the global mean temperature is in question. However, in the actual calculation of the global climate sensitivity and feedbacks, $d \alpha_i/dT$ has never been explicitly computed.  Instead, for each feedback process $i$, one often computes the radiative flux at the top of the atmosphere (TOA) resulting from the change $\Delta \alpha_i$ between the perturbed and unperturbed climates, i.e.,  $\Delta R_i= \partial R/ \partial \alpha_i  \Delta \alpha_i $, and presents the flux normalized by the global mean temperature $\Delta \bar{T}$ as the feedback $\lambda_i$ (with a unit of Wm$^{-2}$ K$^{-1}$). Diagnosing the feedback as such carries an implicit message: $\Delta \alpha_i$ is caused by the global mean temperature change—a perception with no real physical basis. The $\lambda_i$ computed in this manner is often time- and forcing-dependent and, thus does not represent a physical property intrinsic to the climate system.

Given the limitations of the global sensitivity described above, a local feedback perspective for regional climate response and feedback has been developed by reformulating the feedback as a linearization about the local temperature  (e.g., \cite{boer2003aclimate, boer2003bclimate, winton2006amplified, bates2007some, bates2012climate, crook2011spatial, boer2011ratio, kay2012influence, roe2015remote, feldl2013four}): 
\begin{equation}
    \Delta T(\mathbf{x}) = \frac{1}{\lambda_0 (\mathbf{x})} \left[ \Delta R_f(\mathbf{x}) - \Delta (\nabla \cdot F (\mathbf{x})) + \left(\sum_i \lambda_i (\mathbf{x}) \right) \Delta T(\mathbf{x}) \right], 
    \label{eqn:f_local}
\end{equation}
where all variables are spatially-varying (i.e., functions of $\mathbf{x}$), and an energy flux convergence term $(-\nabla \cdot {F(\mathbf{x})})$ must be included to ensure local energy balance. This term can be interpreted as the energy convergence in the atmosphere alone (e.g., \cite{feldl2013four}) or in both ocean and atmosphere (e.g., \cite{hahn2021contributions}), depending on the model configuration. In addition, the energy storage term must be also accounted for if the system is not in equilibrium (e.g., \cite{donohoe2020partitioning, huang2021quantifying}). Now that the surface temperature is a vector, $\lambda_i$ expressed in (\ref{eqn:f_mean}) is a matrix obtained from the product of two Jacobian matrices, $\partial R(\mathbf{x})/\partial \alpha_i (\mathbf{x})$ and $\partial \alpha_i (\mathbf{x})/\partial T(\mathbf{x})$. Provided that all feedbacks are local (i.e., $\lambda_i$ is diagonal with entries $\lambda_i(\mathbf{x})$), equation (\ref{eqn:f_local}) could alternatively be rearranged into a form like equation (\ref{eqn:feedback}):

\begin{equation}
    \Delta T(\mathbf{x}) = \frac{1}{\lambda_0 (\mathbf{x})} \frac{\Delta R_f (\mathbf{x}) - \Delta (\nabla \cdot F)}{\left(1 - \frac{1}{\lambda_0(\mathbf{x})} \sum_i \lambda_i (\mathbf{x})\right)}. 
    \label{eqn:feedback_local}
\end{equation}
where $\lambda_0$ is often expressed as a diagonal matrix under the assumption that Planck feedback is predominantly local. This regional feedback framework has been argued to be more physically intuitive and more advantageous in explaining the pattern of the response in terms of how spatial patterns of forcing and feedbacks contribute to the pattern of the response (\cite{feldl2013four}). It has been used to explain the time variation of the effective global climate sensitivity in the transient climate adjustment to the increase of CO$_2$ forcing with the evolution of the pattern of the surface warming and to resolve the conundrum of the different climate sensitivities between climate models and historical observations—the so-called “pattern effect” (e.g., \cite{armour2013time, rose2016effects, zhou2016impact, proistosescu2017slow, andrews2018dependence, dong2019attributing, dong2020intermodel, rugenstein2020equilibrium, lin2021dominant, bloch2021climate, forster2021earth}). However, in all the existing applications of the regional feedback analysis (e.g., \cite{armour2013time, dong2019attributing, dong2020intermodel}), the locality assumption has been made for $\lambda_i (\mathbf{x})$ and this is equivalent to assuming that the kernel Jacobian matrix $ \partial R(\mathbf{x})/ \partial \alpha_i (\mathbf{x})$ and the sensitivity matrix $ \partial \alpha_i (\mathbf{x})/ \partial T(\mathbf{x})$ are both diagonal. This assumption is ad hoc and unphysical, disregarding the fundamental interconnectedness of the climate system. At the process level, the change $\Delta \alpha_i$ is not locally driven by the surface temperature but governed by the complex nonlocal interactions between the physics and dynamics of the climate system. Perhaps more consequential in the construct of the local feedback framework is that the energy transport in the ocean and atmosphere is often treated as an afterthought, decorrelated from the radiative feedback processes (e.g., \cite{lu2009new, pithan2014arctic, liu2018sensitivity, hahn2021contributions}). Overlooking the covariation between the circulation and the physical feedbacks in the climate system has been the crux of the existing feedback analysis frameworks for studying climate sensitivity (\cite{zhang1994diagnostic, colman1997non, huang2021nonlinear}). As a result, these feedback analysis frameworks reviewed above are overall diagnostic in nature and lack predictive power to predict the response to a given forcing and vice versa.

Moreover, climate sensitivity about the global mean surface temperature falls short of adequately informing climate policy-making and adaptation planning, which operates across various scales from local to global. More relevant to reliable regional climate information may be a predictive understanding of the generalized climate sensitivity concerning the response of not only the mean but also the pattern of global surface temperature, precipitation, sea level, wind, etc. Currently, despite multiple phases of coupled model intercomparisons (CMIPs), regional climate projections still suffer from large uncertainties and lack model consensus (e.g., \cite{neelin2006tropical, knutti2013robustness, shepherd2014atmospheric, xie2015towards}). There is an urgent need for a pattern-aware analysis framework to understand the forcing and feedback mechanisms underlying the emerging patterns of climate response and to provide robust and reliable regional climate projections to users of climate information.
This study represents the first attempt towards this goal by devising a nonlocal, pattern-aware climate response and feedback analysis framework. As a first attempt, the effort focuses only on the temperature response and is equivalent to extending the 0-dimensional framework of (\ref{eqn:feedback}) to a finite-dimensional one, based on the same local energy balance as described in equation (\ref{eqn:f_local}), but with the energy fluxes in the ocean and atmosphere parameterized in a simple but nonlocal form. Unlike conventional frameworks, our approach simultaneously estimates parameters for oceanic and atmospheric energy transport, along with physical feedback parameters, using a data-driven method. This allows for capturing the universal response and feedback relationships within the climate system, organized into patterns that provide both interpretability and potential generalizability. 
Our approach is guided by a vision afforded by the dynamical system view of the climate: despite the very high dimensionality of the climate state space, the slow-evolving response can often be captured by much lower-dimensional attractors due to the dissipative and self-organized nature of the system dynamics of the climate (\cite{franzke2005low, majda2008applied, lucarini2016extremes, faranda2019attractor, hummel2023reduction}). As it turns out, estimating the sensitivity and feedback matrices and the parameters therein involves solving an ill-posed inverse problem of a climate system, which requires regularized, parsimonious inversion in a reduced space. 

Our efforts result in a reduced-order representation of the climate system obtained by truncating the learned forcing-response-feedback relationship, which exhibits encouraging skill in predicting the response to any forcing pattern and modest skill in finding the optimal forcing for any physically realizable temperature pattern with potential for climate intervention applications. Representing and deciphering the feedbacks in reduced space (i.e., truncated SVDs) also serves as an interpretation framework for the forced response in terms of the superposition of the dominant SVD modes. The robustness of these modes may potentially qualify them as dynamic fingerprints for detecting robust regional climate response.

The remainder of the paper is structured as follows. Section 2 introduces a conceptual framework based on column energy balance for regional response and feedback. This is followed by an introduction of a set of Green's function experiments and a data-driven method to fit a comprehensive linear response function (CLRF) using the experimental data. The learned CLRF encompasses the feedback information in the target climate system. In Section 4, we examine the patterned feedbacks by the SVD of the CLRF. The first SVD mode turns out to correspond to global warming with Arctic amplification, and our pattern-aware feedback analysis sheds new light on the deterministic mechanisms and the associated feedbacks of the Arctic amplification (AA). We then examine the robustness of the trained CLRF to the internal noise of the system. The skill of the trained CLRF is then demonstrated through the prediction of an independent test case that is not seen during the training. The relative robustness and the promising skill of the CLRF provide us some confidence in this first proof-of-concept effort toward a systematic, patterned analysis framework for climate response and climate forcing. The paper concludes with a discussion of the possible next steps to develop the effort towards producing more useful, robust, and relevant patterned climate information.

\section{A patterned feedback framework}
\begin{figure}[h]
    \centering    \includegraphics[width=\textwidth]{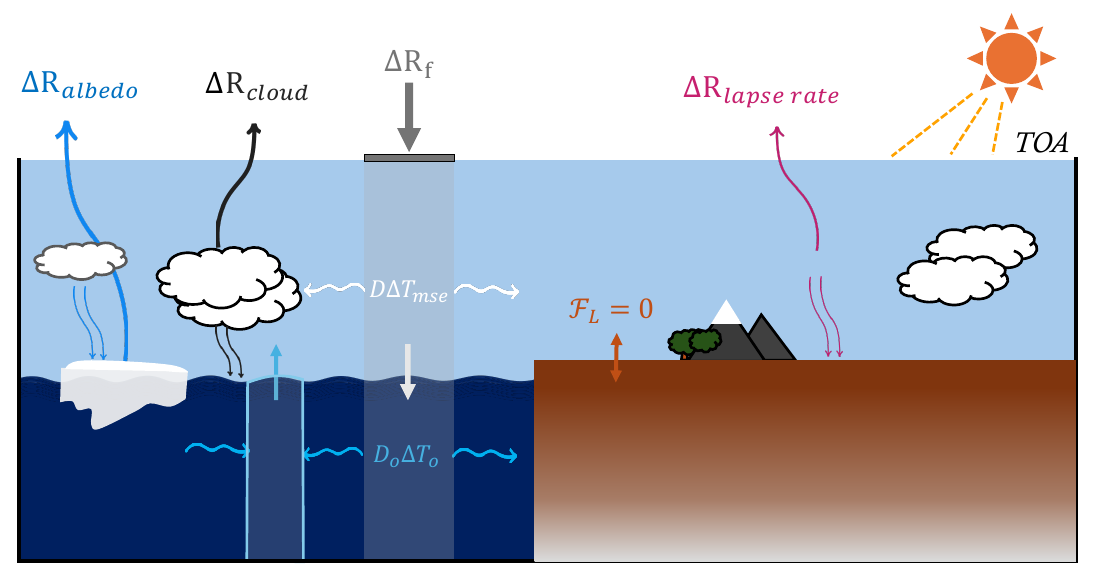}
    \caption{Schematic of feedback framework. Upon a downward solar radiation perturbation, a coupled climate system responds with not only changes in the atmospheric and oceanic heat transport but also radiative adjustments at both the top of the atmosphere and the air-sea interface. Three atmospheric feedback processes are considered in the construct of the feedback framework: surface albedo, cloud, and lower-tropospheric lapse rate, which are represented by feedback variables $\alpha_{alb}, \alpha_{cod}, \alpha_{lr850}$, respectively. }
    
    \label{fig:feedback-scheme}
\end{figure}
In order to investigate the equilibrium response of the climate to external radiative forcing, we consider the local energy balance from a column perspective as shown in Fig. \ref{fig:feedback-scheme}. Any external spatially-varying radiative forcing, $\Delta R_f(\mathbf{x})$ at location $\mathbf{x}$, assuming quasi-equilibrium, must be balanced by the atmospheric and ocean energy flux divergence/convergence $\Delta (\nabla \cdot F(\mathbf{x}))$, and local and remote radiative feedbacks $R(\mathbf{x})$. The climate system we consider in this study is assumed to have equilibrated over land, i.e., there is no net energy transfer across the land surface ($\mathcal{F}_L(\mathbf{x}) = 0$). Thus, we consider an energy balance framework from a column perspective for the linearized equilibrium response that accounts for changes in atmospheric and oceanic transport and the associated nonlocal feedfbacks as follows.


\begin{equation}
        \Delta R_f(\mathbf{x}) = \mathbf{D} \Delta T_{mse}
        (\mathbf{x}) + \mathbf{D}_o \Delta T_o(\mathbf{x}) - \sum_{i} \frac{\partial R}{\partial \alpha_i}  \Delta {\alpha_i}(\mathbf{x}), 
        \label{eqn:energy_balance_nl}
    \end{equation}
In this formulation, $\Delta T_{mse}$ is the baseline response variable, representing the linear response of the density-weighted vertical integration of the moist static energy (MSE) with a temperature unit ($T_{mse}=MSE/c_p$). With this choice of the baseline variable, the radiative effects of column temperature and water vapor are accounted for in a single variable, and there is no need to consider phase change and the related diabatic heating (condensation and fusion) in the atmosphere in the construct of the climate feedback framework. This choice also accounts partially for the cancellation between water vapor and lapse-rate feedbacks (Held and Shell, 2012).  $\Delta T_o$ indicates the linear response of sea surface temperature (SST), and the second term is intended to capture the feedbacks proportional to the SST changes.
$\Delta \alpha_i$ represents the linear changes in three feedback variables: surface albedo, clouds, and lapse rate. The choices of these three feedbacks are based on the existing understanding of the dominant radiative feedbacks (other than the baseline feedbacks) under climate warming (e.g., \cite{bony2005marine, ceppi2017cloud, held2000water, holland2003polar, klein1993seasonal, soden2006assessment, screen2010central,previdi2021arctic,zhang2018local}).

As a result of the choices of the baseline variable and feedback variables, the first term on the right-hand side of the equation accounts not only for the TOA+surface radiative change due to $\Delta T_{mse}$ but also for the `diffusive' effects of the mean atmospheric circulation on the $T_{mse}$ anomalies due to the atmospheric circulation. Matrix $\mathbf{D}$ will be referred to loosely as diffusion kernel here. The same treatment is applied to the SST-related feedbacks. The second term $\mathbf{D}_o \Delta T_o(\mathbf{x})$ is intended to parameterize the effect of the oceanic feedback, as SST is the only ocean state variable in the construct of the feedback framework. This term is dominated by the ocean heat uptake (OHU) response, which, on the time scale when the ocean mixed layer is close to equilibrium, is predominantly balanced by the heat divergence/convergence in the ocean mixed layer. Therefore, $\mathbf{D}_o$ can be interpreted as the diffusivity matrix for the ocean heat energy transport. For both ocean and atmosphere, the nonlocal, off-diagonal components of the diffusion matrix allows us to capture the teleconnection effect in the climate system.
Jacobian matrix $\frac{\partial R}{\partial \alpha_i}$ describes both local and non-local radiative feedbacks due to local changes in the feedback variable $\alpha_i$. Since $\frac{\partial R}{\partial \alpha_i}$ is nondiagonal, it captures the radiative contributions to teleconnections. Note that the column perspective is taken here, the radiative feedback from certain variable $\alpha_i$ represents the net feedback through both the TOA and surface radiative fluxes, which are indicated by the upward and downward arrows, respectively, in Fig. \ref{fig:feedback-scheme}. The key distinction of the feedback analysis framework here from the conventional approaches is its holism that accounts for the inseparability of dynamical and radiative feedbacks and the teleconnectedness inherent in the climate system. 


Expressing (\ref{eqn:energy_balance_nl}) in matrix form, we have: 
\begin{align}
    \Delta R_f &= \begin{bmatrix}
     \mathbf{D} & \mathbf{D}_o & -\mathbf{K}_{alb} &  -\mathbf{K}_{cod} &  -\mathbf{K}_{lr}
 \end{bmatrix} \begin{bmatrix}
\Delta T_{mse} \\ \Delta T_o  \\\Delta \alpha_{alb} \\ \Delta \alpha_{cod} \\ \Delta \alpha_{lr}
\end{bmatrix}, 
\label{eqn:ls}
\end{align}
which can be solved as a multivariate regression problem to fit for the comprehensive kernel, $$\mathcal{K} = \begin{bmatrix}
     \mathbf{D} & \mathbf{D}_o  & -\mathbf{K}_{alb} &  -\mathbf{K}_{cod} &  -\mathbf{K}_{lr}
 \end{bmatrix}, $$ 
 given the linear responses stacked in the right square brackets. The linear responses are those from Green's function experiments using a coupled climate system, which will be described in the next section. The sign convention is such that a positive term on the right hand side of (\ref{eqn:ls}) represents an energy loss in the column energy balance. Framed as such, equation (\ref{eqn:ls}) can be thought of as a surrogate model and $\mathcal{K}$ as a comprehensive linear response function (CLRF) for the original climate system,
\begin{equation}
     \Delta R_f = \mathcal{K} \Delta X.
\label{eqn:clrf}
 \end{equation}
where $\Delta X = [\Delta T_{mse} \quad \Delta T_o \quad \Delta \alpha_{alb} \quad \Delta \alpha_{cod} \quad \Delta \alpha_{lr}]^T$. Here, $\Delta \alpha_{alb}$ is the percentage change in the annual mean surface albedo, $\Delta \alpha_{cod}$ represents the response in the annual mean effective cloud optical depth (COD), parameterized as the product of the cloud fraction and the total cloud water path (COD = 1.5 $\times 10^{-2}$  $m^2 Kg^{-1} \times CWP \times$ cloud fraction), assuming a representative cloud droplet size, and $\Delta \alpha_{lr}$ is taken to be the difference between the surface temperature and the temperature at 850hPa. Studies have found a strong anticorrelation and offsets between the lapse rate and water vapor feedbacks in the tropics and midlatitudes (\citet{colman2021water}), which has been partially captured by $\Delta T_{mse}$. Therefore, a near-surface representation of the lapse rate is purposed to focus on the high-latitude contributions due to near-surface inversion feedback. To emphasize this difference, we use $\Delta \alpha_{lr850}$ for the response and $\mathbf{K}_{lr850}$ for the corresponding kernel to denote this term in the remainder of the paper. 

Furthermore, the matrix-form feedback associated with the variable $\alpha_i$ in this framework can be easily constructed as follows,

\begin{equation}
    \mathbf{\Lambda}_{\alpha_i} = \mathbf{K}_{\alpha_i} \frac{\partial {\alpha_i}}{\partial T_{mse}},
    \label{eqn:Lambda-def}
\end{equation}
such that,

\begin{equation}
   \Delta {\alpha_i} (\mathbf{x}) =  \frac{\partial {\alpha_i}}{\partial T_{mse}} \Delta T_{mse} (\mathbf{x}).
   \label{eqn: T-cond-response}
\end{equation}
The feedback $\mathbf{\Lambda}$ defined here is a tensor that accounts for the nonlocal effects (compare with (\ref{eqn:f_local})) and the corresponding feedback contribution to a patterned temperature response $\Delta T_{mse}(\mathbf{x})$ can then be computed as 

\begin{equation}
  \Delta R_{\alpha_i} (\mathbf{x}) = \mathbf{\Lambda}_{\alpha_i} \; \Delta T_{mse}(\mathbf{x}).
  \label{eqn:F-decomposition}
\end{equation}

 Given (i) the unique choice of the set of baseline and feedback variables, (ii) the column feedback perspective, and (iii) the holistic data-driven approach to estimate the diffusivity and radiative kernels simultaneously, it may not be straightforward to compare the resulting feedback fluxes (in $W m^{-2}$) for a climate change scenario directly with conventional feedback analysis.    

\section{Methods}
\subsection*{Green's Function Experiments}
\begin{figure}[h]
    \centering
    \includegraphics[width=\textwidth]{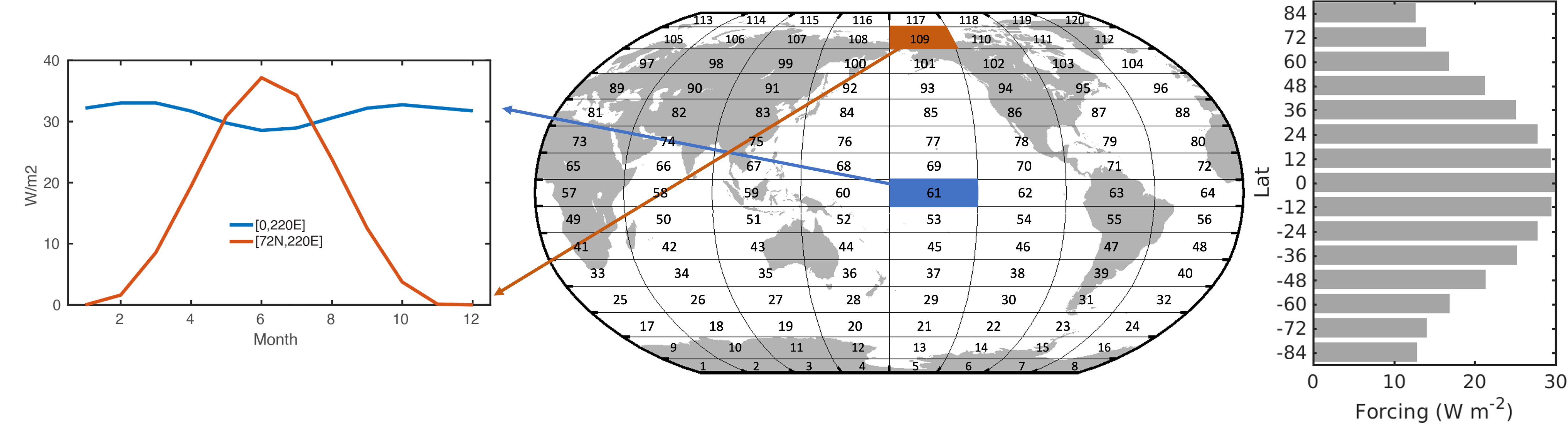}    
    \caption{The array of the forcing perturbation patches for the Greens function experiments and the latitudinal dependence of the patch forcing on the latitude.}
    \label{fig:gf-scheme}
\end{figure}

The data set used to train the linear system in (\ref{eqn:clrf}) and thus evaluate the comprehensive kernel $\mathcal{K}$ is a suite of Green's function experiments consisting of 120 pairs of patch experiments. These experiments are performed with the Community Earth System Model version 1 (CESM1) with CAM4 as its atmospheric component from the National Center for Atmospheric Research (NCAR). For each patch, a pair of shortwave radiation perturbations ($\approx 30$ W/m$^2$), with the same magnitude but opposite sign, is applied at TOA over a rectangular area of size 45$^\circ$ $\times$ 12$^\circ$. This is achieved by scaling up and down the solar constant by a small fraction, respectively, such that the corresponding annual mean perturbation at the equator reaches 30 W/m$^2$. The actual annual mean forcing perturbations as a function of latitude are shown in the right-hand side panel of Fig. 2. These perturbation simulations branch out from the 401st year of the control run and are integrated for 100 years. Only the last 50 years are used for analysis, with the prior years discarded as spin-up. The 50-year averaged response is meant to represent the quasi-equilibrium response of the CESM1 climate system, but the general characteristics are not expected to differ qualitatively from their equilibrium counterparts (e.g., \cite{huang2017pattern}). More details on these experiments are documented in \citet{liu2022neutral}. 

From each pair of patch experiments, the linear response vector $\Delta X$ (up to second order accuracy) associated with a given patch forcing can be obtained as $\Delta X = (\delta X^+ - \delta X^-$)/2, where $\delta X^+$ and $\delta X^-$ are the responses to the positive and negative forcing, respectively. Thus, a set of 120 independent cases is obtained, which is then used to evaluate $\mathcal{K}$. Note that the forcing used in (7) and (8) to train $\mathcal{K}$ should be the effective radiative forcing, that is, the shortwave radiation perturbation times one minus the planetary albedo.

For validation, a set of experiments with a globally applied TOA shortwave forcing has been performed. This forcing has the same spatial distribution as the aggregation of all the 120 patch forcings but with a magnitude of 1/5th as much. This case will be referred to hereafter as the global 6 W/m$^2$ case hereafter as the forcing peaks at 6 W/m$^2$ at the equator. Like the patch experiments, both positive forcing and negative forcing cases are performed, and only the linear component of the response during the last 50 years of the 100-year integration is used for analysis. This extra set of runs serves as an independent test case to evaluate how well the diffusion and radiative kernels learned from the Green's function experiment can predict the response to the global 6 W/m$^2$ forcing and predict the forcing from the given response of this test case. 
\subsection*{Kernel evaluation}
The coarseness in the design of the Green's function forcing patches manifests as locally patchy responses in the training data and therefore imposes a natural constraint on the resolution at which the forcing-response relationship can be studied. To reflect this, the response and forcing functions are expressed in terms of an orthogonal basis consisting of 120 non-overlapping patches, the same as the forcing patches used for the Green's function experiments but suitably normalized as follows. For example, the $j^{th}$ patch is given by,

\begin{equation}
    \phi_j(\mathbf{x}) = \frac{1}{\max(F_j(\mathbf{x}))} F_j(\mathbf{x})
    \label{eqn:basis-def}
\end{equation}
where $F_j(\mathbf{x})$ is the annual mean of the applied effective forcing on patch $j$. Now, the linear equilibrium response $\Delta T$ to any external forcing can be projected onto the patch basis and expanded as,

\begin{equation}
    \Delta T \approx \sum_j \Delta T_j^{\phi} \phi_j(\mathbf{x}). 
    \label{eqn:proj-to-basis}
\end{equation}
This can be done for all the response variables and the corresponding coefficients can be obtained similarly. Note that the projection of the forcing $F_j$ for the Green's function experiment at patch $j$ results in a vector of length $120$ with a non-zero entry at the $j^{th}$ index. Therefore, $\Delta R_f$ in (\ref{eqn:ls}) when expressed in this basis, results in a diagonal 120 by 120 matrix. Once the coefficients for all the terms in (\ref{eqn:ls}) are obtained, a reduced linear system in terms of the patch basis coefficients can be obtained as,

\begin{align}
    \Delta R_f^{\phi} &= \begin{bmatrix}
     \mathbf{D}^{\phi} & \mathbf{D}_o^{\phi} & -\mathbf{K}_{alb}^{\phi} &  -\mathbf{K}_{cod}^{\phi} &  -\mathbf{K}_{lr850}^{\phi}
 \end{bmatrix} \begin{bmatrix}
\Delta T_{mse}^{\phi} \\ \Delta T_o^{\phi} \\\Delta \alpha_{alb}^{\phi} \\ \Delta \alpha_{cod}^{\phi} \\ \Delta \alpha_{lr850}^{\phi}
\end{bmatrix}.
\label{eqn:phi-ls}
\end{align}
One could solve (\ref{eqn:phi-ls}) for $\mathbf{D}^{\phi}$, $\mathbf{D}_o^{\phi}$ and the feedback kernels $\mathbf{K}_{\alpha}^{\phi}$ by using a pseudoinverse which provides the least-squares solution. In the actual inversion, the response variables are nondimensionalized by the corresponding local maximum. For convenience, the superscript $\phi$ will be omitted in the following discussion of kernel matrices.

The multi-regression problem described above is an intrinsically underdetermined linear system due to multi-collinearity: there is a strong linear dependence of the feedback variables on the temperature response variable. In particular, each feedback variable $\Delta \alpha$ can be reconstructed linearly from the temperature response as  $\Delta \alpha = \frac{\partial \alpha}{\partial T} \Delta T$. As such, $\mathcal{K}$ is inherently rank-deficient, and this issue cannot be overcome with more data. We use regularization, specifically ridge regression, to solve this multicollinear inverse problem. The solution from ridge regression minimizes the global sum of the squared energy budget residual ($\epsilon^2 = \| \Delta R_f - \mathcal K \Delta X\|^2$) and a penalty term ($a \sum K_{ij}^2$) which allows for noise filtering \citep{james2013introduction}. The penalty term has a tuning parameter $a$ and this parameter is optimized based on the skill of the model in the reconstruction of an independent test case. The resultant $\mathcal K$, which we will refer to as the comprehensive linear response function (CLRF), therefore provides a physically constrained feedback decomposition framework through enforcing patch-wise energy balance.


The system captured by (\ref{eqn:phi-ls}) represents a reduced representation of the CESM climate system in several senses. First, the radiative feedbacks has been reduced to those related to only three feedback variables, and the feedback-induced fluxes are lumped together as a net contribution to the column budget, whether from the TOA or the surface. The radiative fluxes between longwave and shortwave are also not distinguished. The contribution to the energy balance from the atmospheric and ocean circulation has been simply parameterized in a linear form. Second, all the energy balance terms and the associated kernel matrices have been projected on the forcing patch basis, amounting to coarse-graining the system from the size of the model grids to the size of the forcing patches. Last, the resultant $\mathcal{K}$ matrix can be further decomposed into orthogonal SVDs, which can be truncated further. As will be shown later, the truncated system represents the more robust component of the original climate system than the full system (\ref{eqn:phi-ls}). It must also be stressed that the learned radiative feedback is from an atmospheric column perspective and should be distinguished from the TOA perspective commonly used in the climate sensitivity literature. For example, the well-known positive TOA lapse rate feedback contributing to Arctic amplification under global warming becomes strongly negative when the lapse rate feedback to the surface radiative flux is considered as a part of the column budget, due to the sharp increase of the temperature kernel (for the downward longwave flux) with decreasing height near the surface (see Fig. 4d in Huang and Huang 2023).

\section{Results}
\subsection*{4.1 Kernel structure}

\begin{figure}[h]
    \centering
    \includegraphics[width=0.7\textwidth]{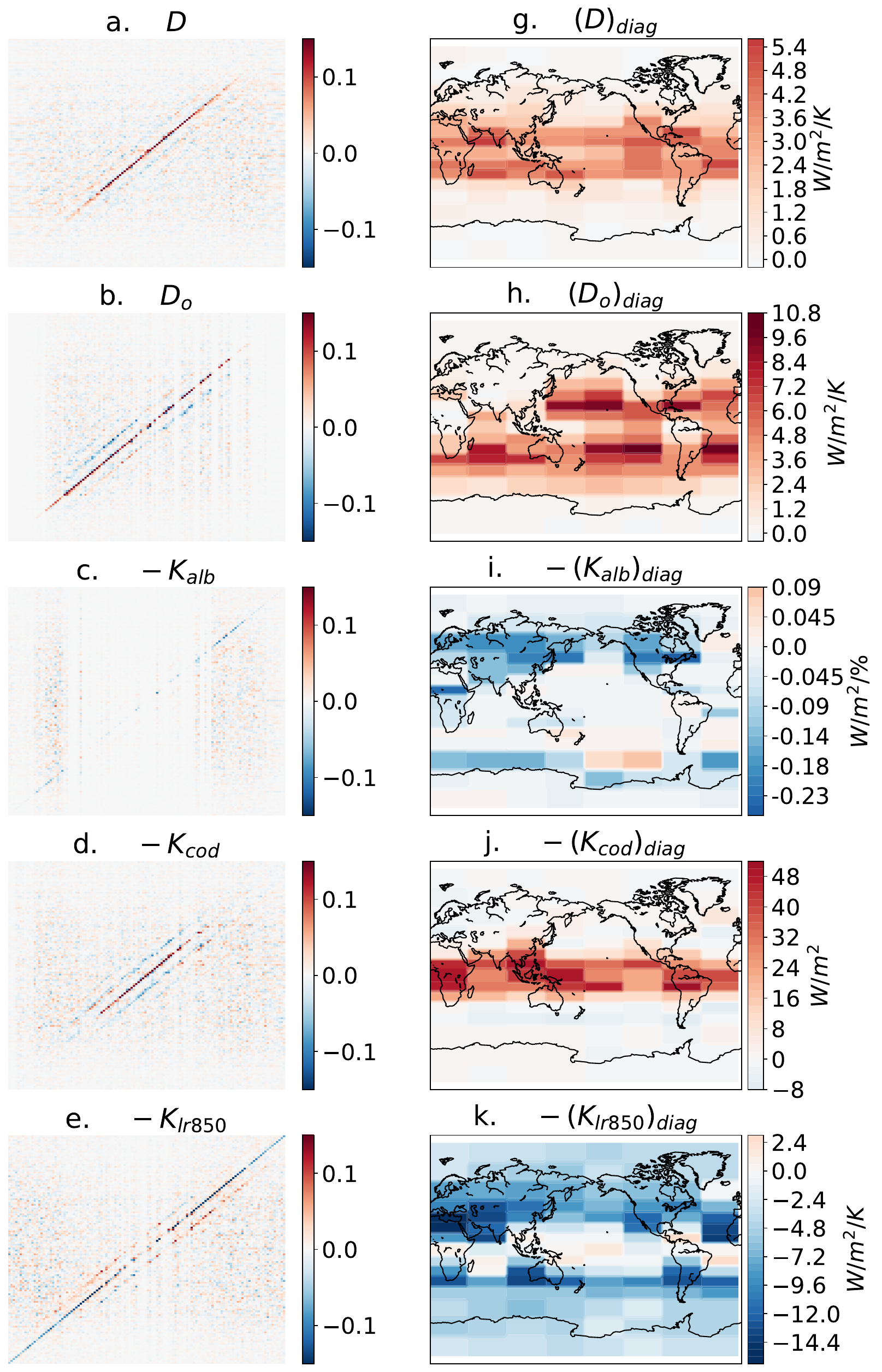}    
    \caption{Results of the estimated CLRF. (a-e) The matrix values of the individual kernels constitute the CLRF. (f-j) The dimensionalized diagonal (local) component of the kernels. Note that COD is a unitless quantity.}
    \label{fig:k-struct}
\end{figure}

Fig. \ref{fig:k-struct} shows the learned diffusion and radiative kernels of the CLRF. The diagonal component, representing the local feedback, tends to be dominant in all cases, while the off-diagonal components are also evident in each case: say, entry $(i,j, i \neq j)$ of $\mathbf{K}_{alb}$ describes the nonlocal feedback from the albedo change at patch j to the net radiative flux change at patch i. The multi-diagonal structures in the diffusion kernels are consistent with what would be expected from discretizing a second-order diffusion operator and thus can be thought of as a manifestation of the diffusive nature of the energy dispersion in the atmosphere and ocean. In addition, the diffusive feature appears to influence the cloud and the lapse rate feedbacks. The remaining off-diagonal components can be attributed to large-scale teleconnections in the dynamically coupled CESM climate system.

The local radiative effects, as represented by the diagonal component of each (dimensionalized) kernel are displayed in the right panels in Fig. \ref{fig:k-struct} and can be qualitatively compared with the established understanding of the related feedback processes in the literature. The local component of $\mathbf{D}$ kernel peaks in the off-equatorial regions and the subtropics. Removing the component due to the radiative feedbacks of column temperature and water vapor (see Fig. A1(b), computed using a radiative kernel assuming 1K uniform column warming and the corresponding increase of specific humidity), the residual may be interpreted as the local "diffusive" component of the $\mathbf{D}$ kernel. Its positive tropics-negative extratropics pattern (Fig. A1(c)) represents a damping (amplifying) effect on the local $T_{mse}$ anomaly in the tropics (extratropics), manifesting an intrinsic tendency of poleward energy transport upon warming (e.g., Budyko, 1969; Held and Suarez, 1974). As will be seen later, this is the most important denominator underlying the Arctic amplification pattern of the leading mode of the system (recall that as per the sign convention, positive D implies a damping effect on the response).

The positive diagonal values of $\mathbf{D}_o$ almost everywhere in the global ocean are consistent with the positive local ocean heat uptake (a cooling to the atmospheric column) that would ensue under a warming forcing. A local negative value of $-\mathbf{K}_{alb}$ implies a loss of energy with a reduction of albedo, which appears to be contradictory to the conventional positive albedo feedback observed under climate warming (e.g., \cite{bony2005marine, pithan2014arctic, roe2015remote}). However, here the $-\mathbf{K}_{alb}$ measures the column radiative feedback from surface albedo, the actual radiative kernel calculation indeed confirms its local negativity (see also \cite{huang2017pattern}). 
On the other hand, an increase of clouds in the tropics acts to cool the atmospheric column, suggesting the dominance of the COD shortwave effect over its longwave effect in this climate model. Since it is difficult to estimate the local COD radiative effect from physics-based approaches directly, here we leave the process-level investigation out of the scope of the current methodology-oriented study.

The local lapse rate feedback also appears to be against the conventional wisdom---the negative local value of $-\mathbf{K}_{lr850}$ implies a cooling effect to the air column when the lower-tropospheric lapse rate increases as the surface warms more than the levels above. A careful examination of the radiative kernel for the typical winter temperature profile indeed confirms the sign of the sensitivity of the net column flux to the lapse rate change: the positive LW radiative gain through the TOA tends to be overcompensated by the loss through the surface under surface-confined warming (or an increase in the near-surface lapse rate).

\subsection*{4.2 Forcing-response singular modes}

As one can see from Fig. 3, non-zero values permeate throughout the off-diagonal element of the learned comprehensive linear response kernel $\mathcal{K}$, and many of them are subject to the noise in the data used to train it. To decipher the teleconnection patterns and the related feedbacks embedded in the CLRF, especially the robust ones, we perform a singular value decomposition (SVD) of matrix $\mathcal{K}$ below. 
\begin{align}
    \mathcal{K} = \mathbf{U} \mathbf{\Sigma} \mathbf{V}^T,
\end{align}
where the $i$-th column of $\mathbf{U}$ and $\mathbf{V}$ correspond to the $i$-th forcing and response mode, respectively; $\mathbf{\Sigma}$ is a diagonal matrix with singular values $\sigma_i$. The reciprocal of the singular value $1/\sigma_i$ is a measure of the efficacy of the corresponding mode forcing in creating the mode responses, as such the most excitable mode is that corresponding to the lowest singular value ($1/\sigma_1 \approx 24$). Note that this mode is well separated from the subsequent modes ($1/\sigma_2 \approx 18$). The higher singular modes represent increasingly inefficient forcing patterns that result in weak climate responses, which, therefore, are more susceptible to noise. The SVD decomposition allows us to filter out the noise-prone component in the CLRF by removing the higher modes in a reduced order reconstruction of the kernel $\mathcal{K}$:

\begin{align}
    \mathcal{K}_{m} =   \mathbf{U}_m \mathbf{\Sigma}_m \mathbf{V}_m^T.
\end{align}
The optimal SVD truncation is set to be $m=50$, which is chosen by conducting a grid search for the minimum reconstruction root-mean-squared error (RMSE) for the prediction of an independent test case (subsection 4.3) over a range of values of the $L_2$ penalty parameter $a$ ($0.01\le a \le 1$). We will use this truncated kernel $\mathcal{K}_{m}$ for our subsequent computations.

\begin{figure}[h]
    \centering
    \includegraphics[width=\textwidth]{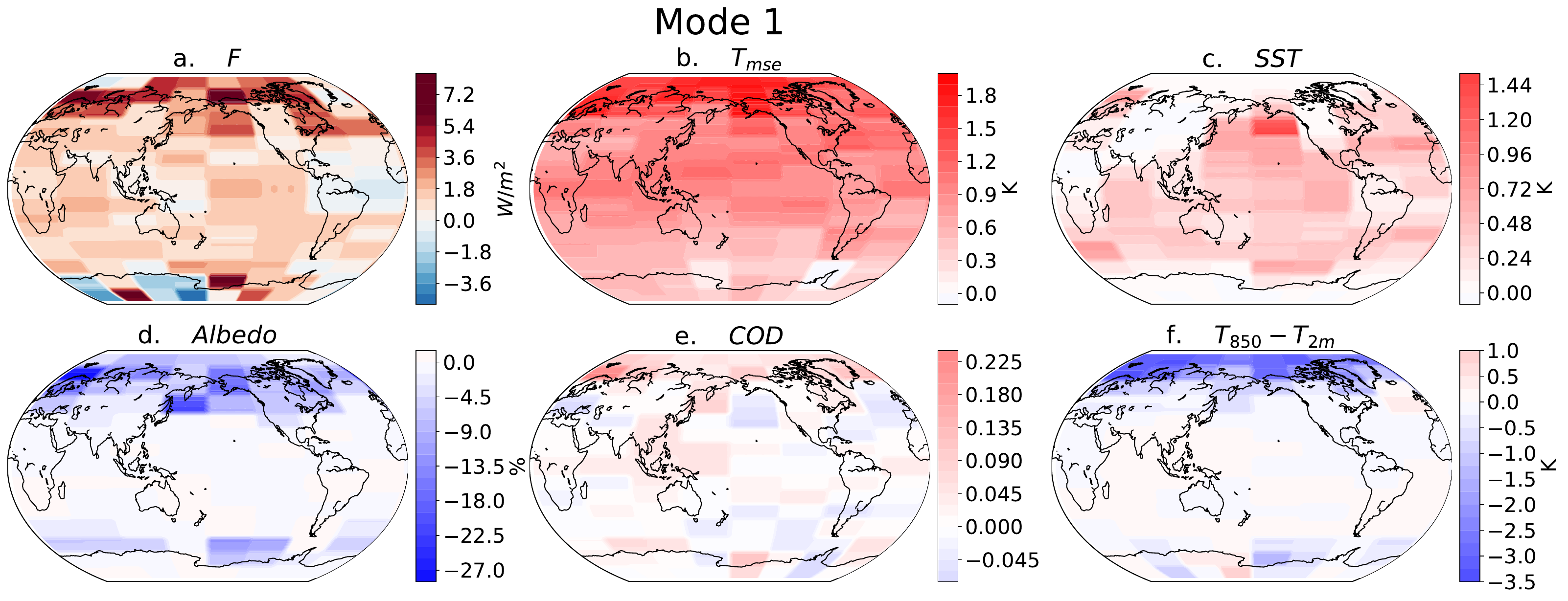}    
    \caption{The forcing-response pair of the first SVD mode of the CLRF.}
    \label{fig:mode-1}
\end{figure}

The dimensional patterns of the forcing, response, and feedback variables of the first singular mode are shown in Fig. \ref{fig:mode-1}. The patterns are obtained from the modes using equations (\ref{eqn:basis-def}) and (\ref{eqn:proj-to-basis}) and by rescaling using appropriate dimensional parameters to revert to dimensional form. 
The $T_{mse}$ pattern features global warming with an amplification in the Northern Hemisphere high latitudes, reminiscent of the Arctic amplification (AA) well established in the climate change research community ( \cite{holland2003polar, alexeev2005polar, dai2019arctic, bintanja2011arctic, england2021recent, previdi2021arctic, hahn2021contributions, taylor2022process}) The corresponding surface temperature warming is even more amplified in the Arctic. Therefore, we can be assured that the leading singular mode captures the AA in this coupled climate system. Physically consistent with the AA warming, SST warms everywhere across the global ocean, showing a positive global ocean heat uptake (Fig. 4c); surface albedo decreases as the climate warms and sea ice and snow melt (Fig. 4d); since the warming tends to be surface-confined in the high latitudes, the lapse rate increases there (Fig. 4f); the corresponding COD change exhibits rather spatially complex structures (Fig. 4e). 
Next, we will investigate how these changes in these feedback variables act in tandem with the $\Delta T_{mse}$ to balance the corresponding TOA forcing (Fig. 4a).

To isolate the feedback contribution for each process, we can multiply the Mode-1 pattern of $\Delta T_{mse}$ with the matrix-form feedbacks $\mathbf{\Lambda}_{\alpha}$:
\begin{equation}
F_{\alpha} (\mathbf{x}) = \mathbf{\Lambda}_{\alpha} \; \Delta T_{mse}(\mathbf{x}).
  \label{eqn:F-decomposition}
\end{equation}
The result is reported in Fig. 5, which shows that globally the Mode-1 forcing is predominantly balanced by the baseline feedback from the $T_{mse}$ warming, with other feedbacks overall making a secondary contribution in their global mean average. The sum of all the feedbacks amounts to a total of $-1.38 W/m^2$, which exceeds the global mean forcing of $1.25 W/m^2$. More careful examination suggests that this inaccuracy stems mainly from the estimate of $\partial{\alpha}/\partial{T_{mse}}$ during the computation of the feedback matrices; the agreement would be much improved were the feedback fluxes computed directly from the kernel $\mathbf{K}_{\alpha}$ and the respective response of the feedback variables. Nevertheless, all the spatial features of the feedback fluxes shown in Fig. 5 still hold at least qualitatively irrespective of the exact approach used to estimate them. Therefore, there should be a physical basis for the patterns of the feedback fluxes. 

Regionally, the direct baseline feedback (Fig. 5a) features a sharp gradient between the northern high latitudes and the remaining region of the NH and the tropics. This can be understood as the result of both the relatively large Planck (cooling) feedback in the lower latitudes compared to the higher latitudes and the tendency of the poleward convergence of MSE through the atmospheric energy transport (as implied by the 'diffusive' component of $\mathbf{D}$ in Fig. A1), while the positive water vapor feedback only works to make $F_{T_{mse}} (\mathbf{x})$ slightly less negative. Nonetheless, $F_{T_{mse}} (\mathbf{x})$ alone cannot balance the Mode-1 forcing (Fig. 5f), it is through all the feedbacks that the system reacts to balance the Mode-1 forcing regionally. In particular, the lapse rate feedback ($F_{lr} (\mathbf{x})$, Fig. 5e) provides the strongest local negative feedback in and around the Arctic to balance the positive forcing there, while the albedo and cloud feedback makes a weak positive contribution within the Arctic circle.

The AA response of Mode-1 can also be understood in terms of its decomposition into the direct $\Delta{T_{mse}}$ response to the Mode-1 forcing plus the components driven by substituting the Mode-1 forcing and feedback variables into the following expression

\begin{figure}[h]
    \centering    
    \includegraphics[width=\textwidth]{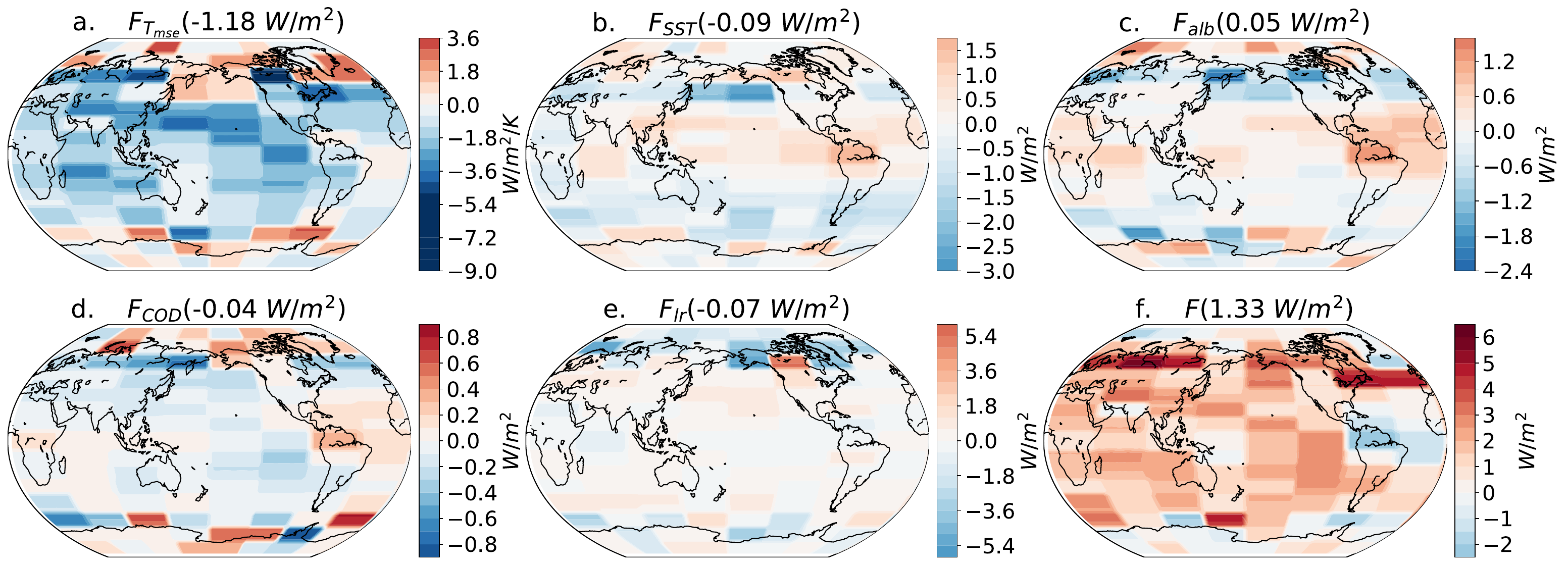} \caption{Feedback contributions to SVD mode 1. Panel (f) is the negative of the sum of the feedback, so the sign can be compared directly with the mode-1 forcing shown in Fig. 4a. The global mean value of each feedback is shown in the title of each panel.}
    \label{fig:lambda_svd1}
\end{figure}

\begin{figure}[h]
    \centering    
    \includegraphics[width=\textwidth]{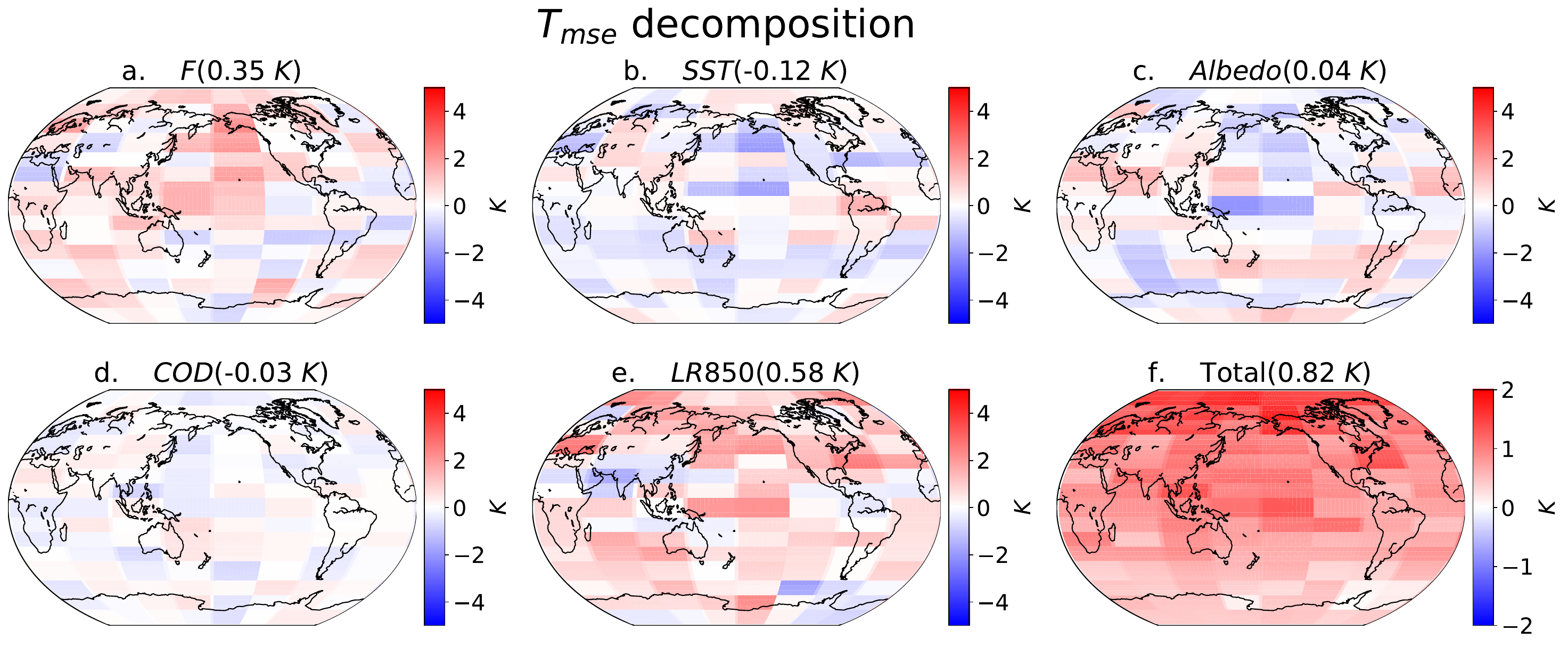} \caption{Attribution of the mode-1 $\Delta T_{mse}$ into the direct response (a) and the feedback responses (b-e). The sum of (a-e) is shown in (f), which should be compared with Fig. 4b.}
    \label{fig:Tmse_svd1}
\end{figure}

\begin{equation}
        \Delta T_{mse}(\mathbf{x}) = \mathbf{D}^{\dag}  \Delta R_f(\mathbf{x})
         - \mathbf{D}^{\dag} \mathbf{D}_o \Delta T_o(\mathbf{x}) + \sum_{i} \mathbf{D}^{\dag}  \mathbf{K}_{\alpha_i}   \Delta {\alpha_i}(\mathbf{x}), 
        \label{eqn:Tmse-decomp}
    \end{equation}
    which is arrived at through a simple manipulation of (6). Here $^{\dag}$ represents the pseudoinverse. The result is shown in Fig. \ref{fig:Tmse_svd1}.   Since the comprehensive kernel is used for the calculation, the decomposition of the global mean $\Delta T_{mse}$ is almost exact, with panels (a-d) combining to accurately reproduce panel (e). Through this decomposition, one can see that global warming is dominated by the direct forcing and the lapse rate feedback, while albedo and cloud feedbacks play a cooling effect globally.  Although the local radiative feedback through lapse rate is negative in the Arctic, working through $\mathbf{D}^{\dagger}$, its contribution to $\Delta T_{mse}$ is positive everywhere around the globe. $F_{SST}(\mathbf{x})$ (Fig. 5b), on the other hand, plays an important role in generating the $\Delta T_{mse}$ gradient that favors AA, via the positive OHU (loss of energy to the ocean) in the mid-latitude oceans.

Collectively, Fig. 4-6 constitutes a self-contained, patterned feedback framework for the mechanisms of the Arctic amplified Mode-1 $T_{mse}$ response. As shown in Fig. 6, the baseline pattern of  $\Delta T_{mse}$ is first set by the direct response (Fig. 6a) to Mode-1 forcing $F$, that is, the would-be response if the forcing is balanced by the MSE transport and the radiative effect of the column MSE, i.e., the effects encompassed in diffusion kernel $\mathbf{D}$. This direct response corresponds to the response component $\frac{\Delta F}{\lambda_0}$ in the 0-dimensional climate feedback problem. Given the interactive nature of the system, the direct response will inevitably induce physically consistent changes in SST, COD, surface albedo and lapse rate (other panels in Fig. 4). Each of these processes can feedback to the original forcing through feedback loop: $\Delta T_{mse}^{dir} \rightarrow \Delta \alpha \rightarrow F_{\alpha} \rightarrow \Delta T_{mse}^{\alpha}$. These feedback fluxes at the TOA and the surface aid in further balancing the forcing, thereby generating additional $\Delta T_{mse}$ that adds to the total Mode-1 $\Delta T_{mse}$. 
The high accuracy in the linear decomposition of the total $\Delta T_{mse}$ (compare Fig. 6f vis-a-vis Fig. 4b) and the physical consistency observed between the feedback variable changes and the corresponding fluxes hold promise for a pattern-aware feedback framework, at least for this most robust aspect of the climate response (i.e., AA). 

The second and third singular modes also show rich spatial structures in both the baseline variable and the feedback variables. Since their relevance to the excitable modes in reality is unclear given the issue of model dependence, we only document their patterns in the Appendix and refrain from further discussion. Both the forcing and response patterns of the SVD modes are orthogonal to each other. Leveraging the orthogonality of the modes, any response to any patterned forcing in the original climate model can be interpreted as the superposition of the SVD modes. The robustness of the SVD modes of the CLRF will be examined next.

\subsection*{4.3 Uncertainty due to Internal Variability}

\begin{figure}[h]
    \centering    \includegraphics[width=0.8\textwidth]{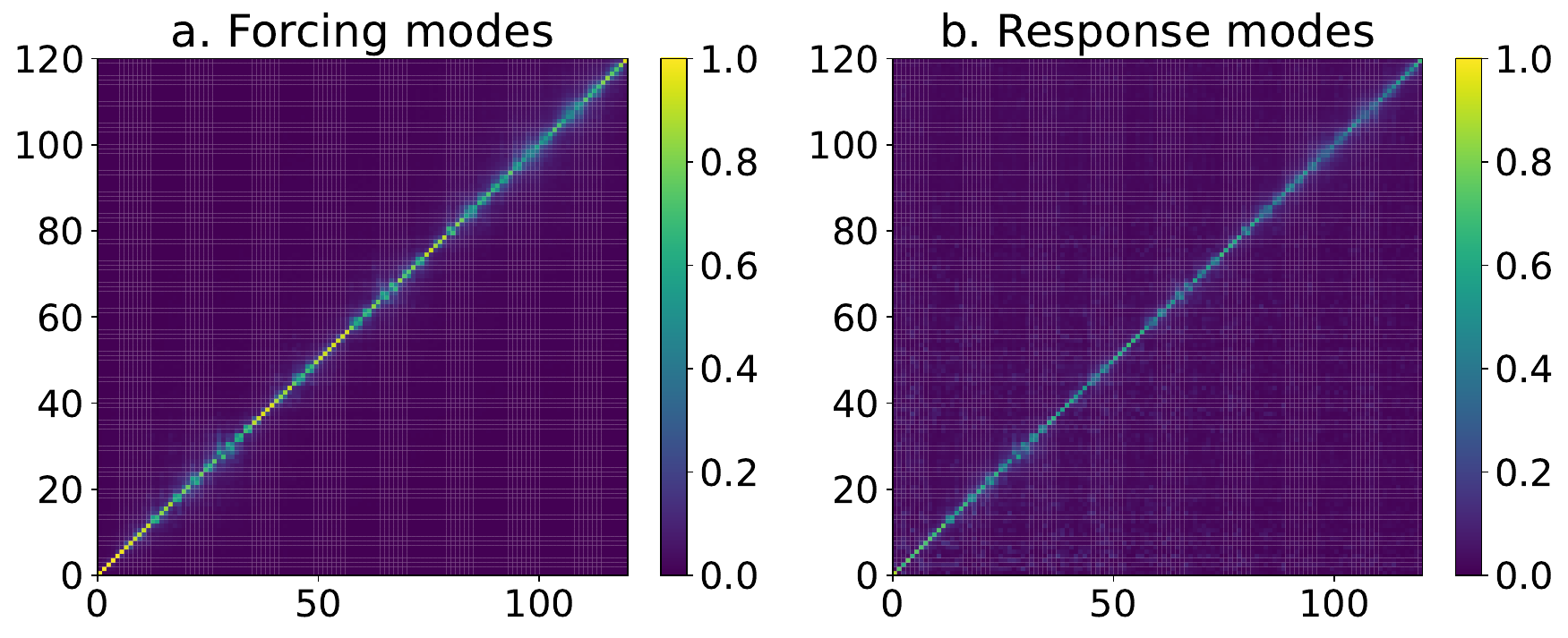}   
    \caption{Uncertainty analysis: 100-realization mean of pattern correlations of forcing modes (a) and response modes (b).}
    \label{fig:pat_corr}
\end{figure}

Although the observables used to train the CLRF are based on a 50-year mean, they are not entirely free of noise, given the significant internal variability of the CESM system across interannual to multidecadal time scales. To assess the impact of the internal noise on the robustness of the learned model and its modes, we first estimate the covariance ($\Sigma (\mathbf{x})$) of the baseline and feedback variables making use of the annual data from the 500 years control run. The resultant covariance matrix $\Sigma (\mathbf{x})$ accounts for spatial co-variation across the baseline and feedback variables. We then generate 100 different realizations of the 120 patch experiments by perturbing the original data for each variable at every location $\mathbf{x}$ using noise independently sampled from a Gaussian distribution with covariance ($\Sigma (\mathbf{x})/50$) -- this is to mimic the variance/covariance in the 50-year mean response arising from internal variability. For each realization, the comprehensive kernel is estimated, and the SVDs thereof are computed. This results in 100 sets of SVDs. For each set, we compute the pattern correlations between each of the pattern of the modes (i.e., $\mathbf{U}$ vectors) with the corresponding pattern of the unperturbed data, and the mean of the 100 sets of correlations are shown in Fig. \ref{fig:pat_corr}a. Similar mean pattern correlation is computed for the response modes ($\mathbf{V}$ vectors) and the result is shown in Fig. \ref{fig:pat_corr}b.
In this matrix-form diagram, large off-diagonal values imply rank jumps of the SVD modes or the sharing of spatial features across modes, implicating the lack of robustness of the modes. The dominance of the diagonal in both matrices indicates the overall robustness of the SVD modes to the internal noise in the response and feedback variables. The diagonal values for the response correlation are overall lower than those for the forcing, which is expected since the responses are directly perturbed when generating the 100 realizations. Given the robustness of the modes and the corresponding CLRF with respect to the internal noise, we next apply the CLRF to an independent test case.





\subsection*{4.4 Prediction of responses and forcing}
\begin{figure}[h]
    \centering    \includegraphics[width=0.8\textwidth]{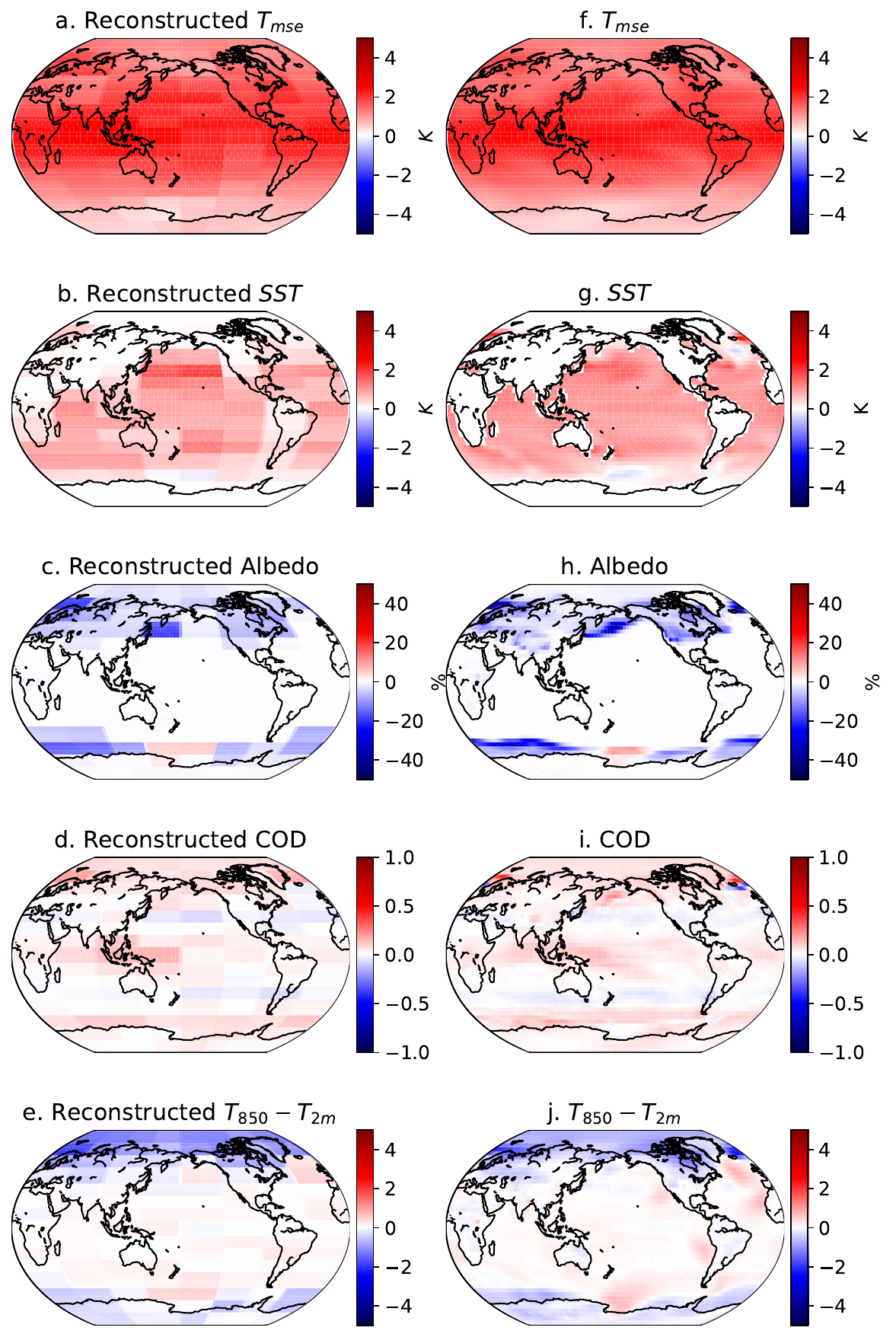}   
    \caption{Response and feedback variables in the global $6 W m^{-2}$ test case in comparison with their reconstruction using the truncated $\mathcal{K}_{m}$.}
    \label{fig:R_recon}
\end{figure}

A real test for the learned CLRF is whether or not it can predict the response to a forcing case it has never seen before, that is, an out-of-sample test. To this end, we use the truncated comprehensive kernel $\mathcal{K}_{m}$ to reconstruct the climate responses for the global $6 W/m^2$ forcing case---a case that is not used during the estimation of CLRF. Note that the actual radiative forcing felt by the coupled climate system when the downward SW radiation is an 'effective' radiative forcing, i.e., the forcing after accounting for the effect of the climatological planetary albedo (note it is different from the ERF often defined in the context of the greenhouse gas forcing). Applying the truncated pseudoinverse of the comprehensive kernel to the effective forcing of the test case ($F_{6W}$), we can predict the response of $T_{mse}$ and the accompanying feedback variables, and the result is presented in Fig. \ref{fig:R_recon}a-e.

\begin{figure}[h]
    \centering
    \includegraphics[width=0.7\textwidth]{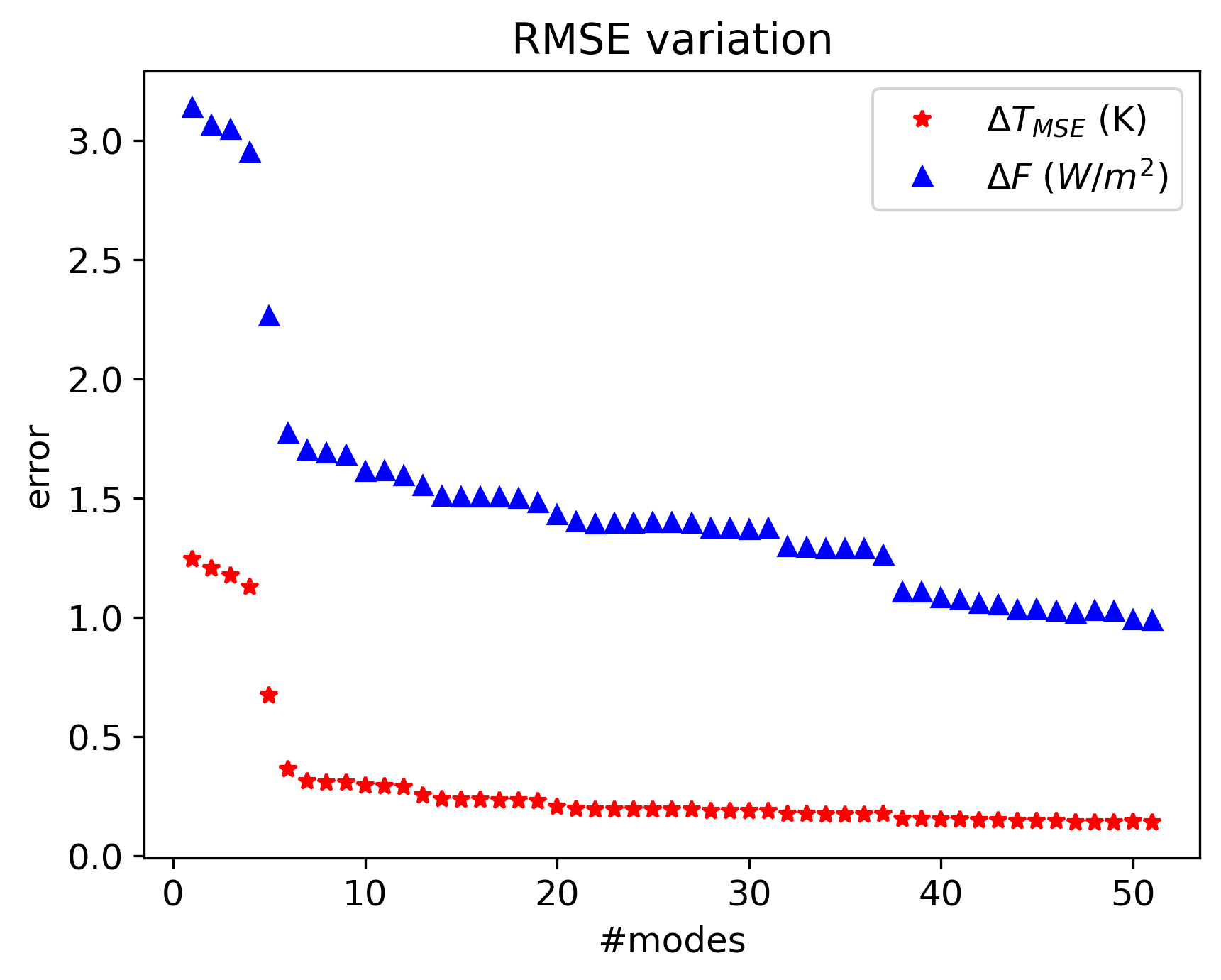}    
    \caption{The dependence of in the root-mean-squared truncation error in forcing and $T_{mse}$ response as a function of the truncation ranks. }
    \label{fig:mse}
\end{figure}

The optimal SVD truncation is set to $m=50$, which is chosen by conducting a grid search for the minimum root mean squared error (RMSE) of the reconstruction over a range of values of the $L_2$ penalty parameter ($a$, $0.01\le a \le 1$). The variation in RMSE with the number of modes for forcing reconstruction and $T_{mse}$ reconstruction using $a_{opt} = 0.5$ is shown in Fig. \ref{fig:mse}.

\clearpage

Compared to the corresponding true response simulated by the CESM model (Fig. \ref{fig:R_recon}f-j), the prediction accurately captures not only the patterns but also the magnitudes for the baseline response and the feedback responses. The physical consistency among the response variables is also evident. It is worth noting that the simulated responses in this test case bear a marked resemblance to those to the forcing of doubling CO$_2$ (not shown), suggesting that the responses for the two different forcing cases are organized by similar modes. Therefore, the patterned feedbacks revealed by the leading SVD modes are likely to be reliable as well. This gives us the confidence to use the truncated feedback matrix to predict the forcing based on the response observed in the test case: 

\begin{align}
    F_{pred, 6W} = \mathbf{\Lambda}_m \Delta T_{mse, 6W}.
\end{align}
where $\mathbf{\Lambda}_m = \mathcal{K}_m \frac{\partial \alpha}{\partial T_{mse}}$. This inverse prediction demonstrates modest skill in capturing the magnitude and even the broad spatial structure of the forcing of the independent test case (compare Figs. 9a and 9b).  To the extent that the predicted forcing can be thought of as a weighted superposition of the optimal forcings of the top 50 modes, the forcing for this test case is effectively optimized, and equation (18) gives the formula for deriving the optimized forcing given a specific $\Delta T_{mse}$ pattern. 

This predicted forcing can, in turn, be decomposed into the baseline feedback to $\Delta T_{mse}$ and the indirect ones through changes in SST, albedo, cloud, and lapse rate (other panels in Fig. 10). 
The direct feedback (Fig. 10c) implies a poleward energy convergence via the atmospheric diffusive transport plus local Planck and water vapor feedbacks, creating the overall meridional structure needed to balance the effective forcing. This baseline pattern is further enhanced by feedbacks from changes in ocean heat uptake ($F_{SST}$), surface albedo, and lapse rate on more regional scales, with cloud feedback playing a secondary role in balancing the effective radiative forcing ($F$).
Regionally, there is considerable cancelation between the OHU feedback and the lapse rate feedback (even more evident in their zonal means) and an anticorrelation between the lapse rate feedback and the albedo feedback, alluding to multicollinearity among them. The predicted OHU feedback ($F_{SST}$) bears a clear spatial resemblance to the net ocean surface heat flux under CO$_2$ doubling forcing reported in \cite{hu2022global}, although in a much coarse-grained form (see Fig. 2 in \cite{hu2022global}). Compared to the pattern of the albedo change itself, the albedo feedback exhibits a more global pattern, epitomizing the non-locality of the feedback. It is worth noting that the zonal mean albedo feedback exhibits a pronounced cooling (to the atmospheric column) near the ice edge in each hemisphere, consistent with the finding of \cite{huang2017pattern}. 
 The multicollinearity among the feedback variables can lead to ambiguity in the interpretation of the individual feedbacks. This might be the main caveat of the pattern-aware analysis framework: holism at the cost of atomism. Therefore, caution is needed when interpreting the specific values of feedback fluxes induced by individual processes, although their joint contribution in balancing the imposed forcing should not be compromised by multicollinearity. 

The local contributions to the radiative feedbacks are also computed using only the diagonal components of the diffusivity and radiative kernels, and the results are displayed in Fig. A4. In comparison with their corresponding full feedbacks, the marked contrast highlights the importance of the non-local, tele-feedbacks, the result of considering dynamics and radiative processes in conjunction. 

\begin{figure}[h]
    \centering    \includegraphics[width=0.46\textwidth]{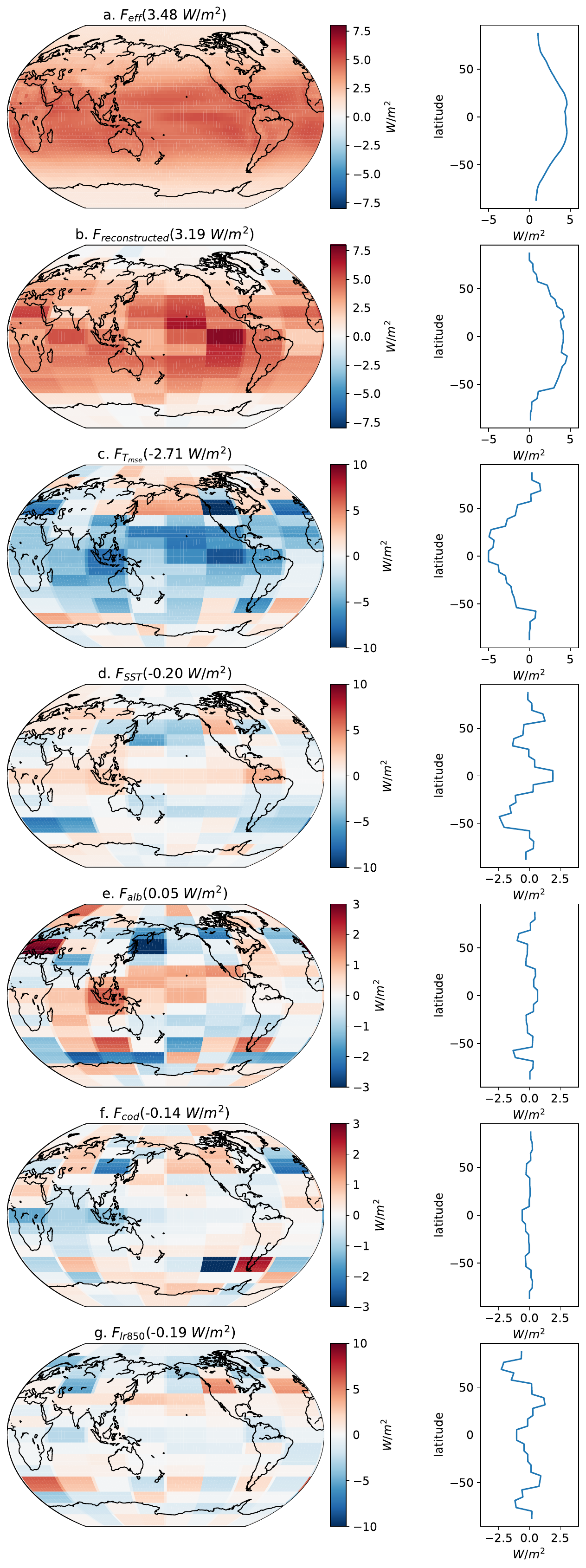}   
    \caption{Forcing reconstruction and decomposition for the global $6 W m^{-2}$ test case. Note the sign convention for the feedback terms is the negative of the RHS of equation (21). The constructed forcing (b) is the negative of the sum of all the feedback terms (c-g). }
    \label{fig:F_recon}
\end{figure}

\clearpage

Given that all feedbacks add together to balance the forcing quantitatively with reasonable accuracy (the relative mean squared error in the forcing reconstruction for the global $6 W/m^2$ test case is $8.5\%$; see also the RMSE error in Fig. \ref{fig:mse}), that is, 

\begin{align}
    F_{6W} \approx F_{pred, 6W} = \mathbf{D}\Delta T_{mse, 6W} + \sum_i \mathbf{\Lambda}^{i}_m \Delta T_{mse, 6W},
\end{align}
 we have achieved closure in the sense that one can retrieve the TOA SW forcing and the associated feedback fluxes for any given $\Delta T_{mse}$ pattern as well as predict both the  $\Delta T_{mse}$ response and the response in the feedback variables for any pattern of the TOA SW forcing.   


%

%

\section{Conclusion and Next Steps}

By training the parameters of a column energy balance equation on a set of Green's function perturbation experiments and using the forcing patches as the basis, we obtain a coarse-grained representation of the climate system in a much reduced space. As such, we have succeeded in constructing a pattern-aware feedback framework in a finite dimension in the same spirit as the 0-dimensional feedback framework used in mainstream climate sensitivity research. 
To the extent that all feedbacks can be added together to balance the forcing, an approximate closure has been achieved, and this can function as a feedback analysis framework applicable to any patterned forcing. Learning feedbacks from data this way is equivalent to solving an ill-posed inverse problem, therefore, careful regularization is essential to achieve physically sensible results. The learned diffusion and radiative kernel matrices are often non-intuitive, due to the covariation among the physical and dynamic processes and the teleconnectedness fundamental to geophysical fluid systems. SVD analysis is conducted to interrogate the learned CLRF and the leading SVD mode reveals that the most excitable pattern of the climate system is a global warming pattern with AA. In addition, the CLRF is applied to an independent test case for both forward and inverse applications, the skills demonstrated from which help further validate the learned feedback relations and build confidence in them. 

The preliminary success of our approach in a perfect-model context opens a promising new avenue for uncovering and understanding emergent patterns in complex climate systems and for developing a metric system to evaluate various Earth system models. For example, the leading neutral mode of the learned CLRF here is an Arctic amplified warming pattern, and our pattern-aware feedback framework reveals its physical underpinning: the basic radiative properties of the moist temperature and the preferred direction of energy transport in the atmosphere. This may be the common denominator that works behind the AA response found in a hierarchy of climate models with different complexity (\cite{merlis2018simple, lu2020neutral, beer2022revisiting, alexeev2005polar}) 

This study represents an initial effort to learn a finite-dimensional representation of the patterned forcing-response relationship. The resulting spatial dimension is quite coarse, with a size of only 120. The next step will be to improve the spatial granularity of the learned forcing-response relationships and the associated feedback patterns by employing advanced dimension reduction techniques. The reduced representation only provides a closure picture in a column-integrated sense, but the actual radiative feedbacks and energy transport occur in a three-dimensional space. There is a need to develop less reduced models to access a more granular, process-level understanding of patterned feedbacks. Achieving this will require more in-depth domain knowledge from climate science and closer collaboration between data scientists and climate scientists.  

This initial effort focuses solely on the linear feedback relationships, while the nonlinear response can be important, as significant nonlinearity has been identified in similar climate systems (e.g., \cite{caballero2013state, jonko2013climate, liu2020sensitivity, lu2020neutral}) 
and the ensuing climate forcing in this century cannot be safely construed to be a small perturbation. A natural extension of the current work is to identify the possible nonlinear feedbacks in the climate system, especially when the forcing magnitude is no longer small. In addition, our study provides a proof-of-concept in a perfect-model context; many of the patterned feedbacks (especially those of the higher modes) are likely to be model-dependent and may improve as the model itself improves. Therefore, it is imperative to extend the analysis to more state-of-the-art climate models. Consensus on the leading modes and the related feedback patterns across different models is also achievable if multiple models exhibit similar patterns and feedbacks for these modes. The emerging patterns with consensus can then serve as the fingerprints (Hasselmann 1979, Hasselmann 1993) for constructing robust regional climate responses. 


\clearpage
\acknowledgments
This work was supported by the Office of Science, U.S. Department of Energy Biological and Environmental Research as part of the Regional and Global Model Analysis program area. The Pacific Northwest National Laboratory (PNNL) is operated for DOE by Battelle Memorial Institute under contract DE-AC05-76RL01830. We thank Malte Stuecker for his comments that helped convey the key message of the work better during the formative stage of the manuscript.  

%
%
\datastatement
The testing and training data used in this study is available here: \begin{verbatim}   https://portal.nersc.gov/cfs/m1199/pkooloth/PAFA/
\end{verbatim}








%



\appendix[A] 

\appendixtitle{Structure of the diagonal of $\textbf{D}$}


\begin{figure}[h]
    \centering    \includegraphics[width=0.9\textwidth]{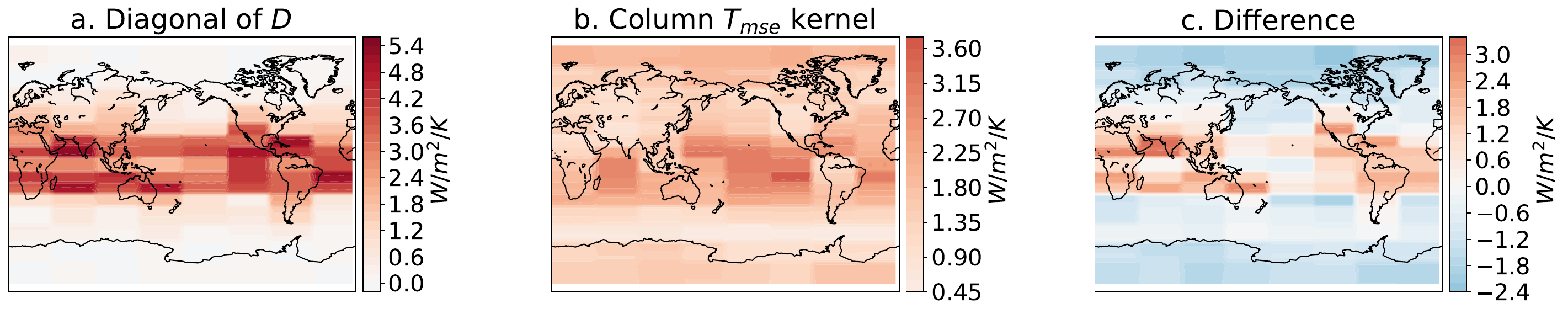}    
    \caption{The diagonal (local) component of diffusion matrix $\textbf{D}$ (a), its radiative component computed using radiative kernel under column uniform 1K warming and consistent moistening (b), and the residual (c) for the 'diffusive' component.}
    \label{fig:Diffusion}
\end{figure}
\clearpage
\appendixtitle{Higher SVD modes of the CLRF}

\begin{figure}[h]
    \centering
    \includegraphics[width=0.9\textwidth]{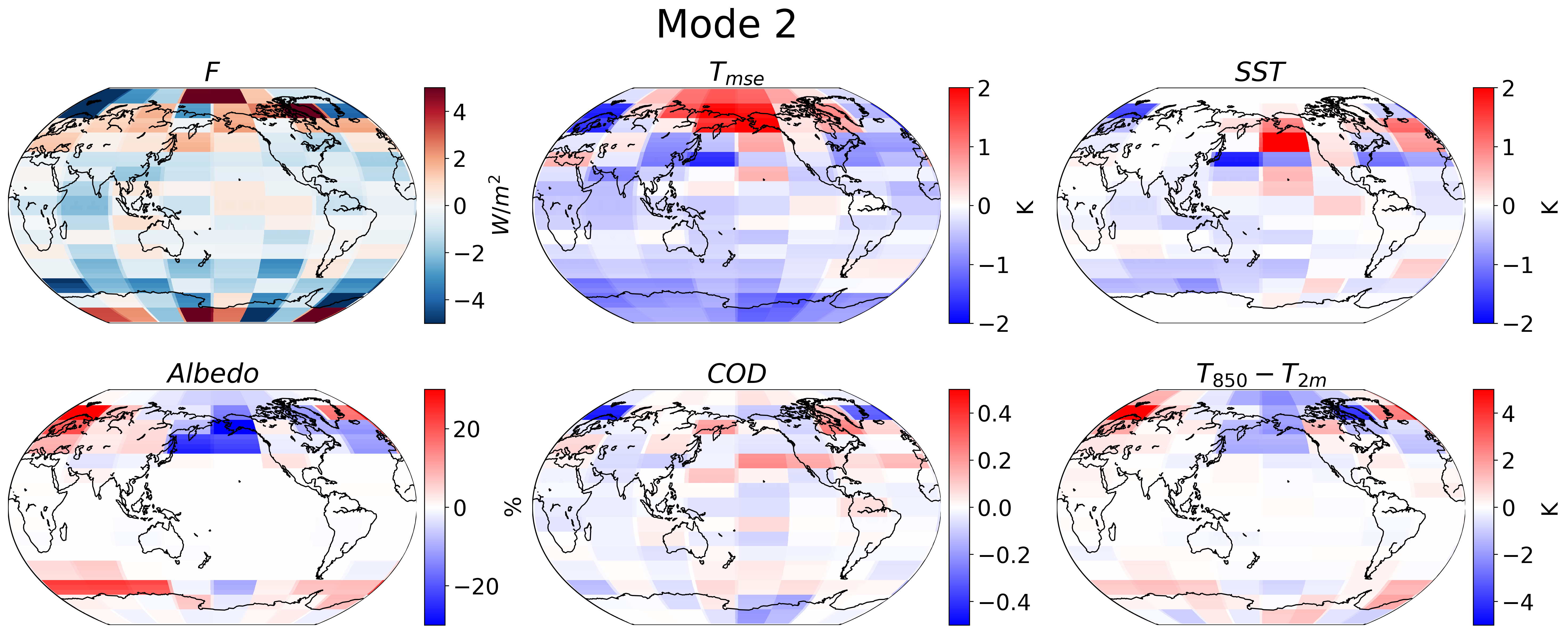}    
    \caption{As Fig. 4, but for the forcing-response pair of SVD mode-2.}
    \label{fig:mode-2}
\end{figure}

\begin{figure}[h]
    \centering    \includegraphics[width=0.9\textwidth]{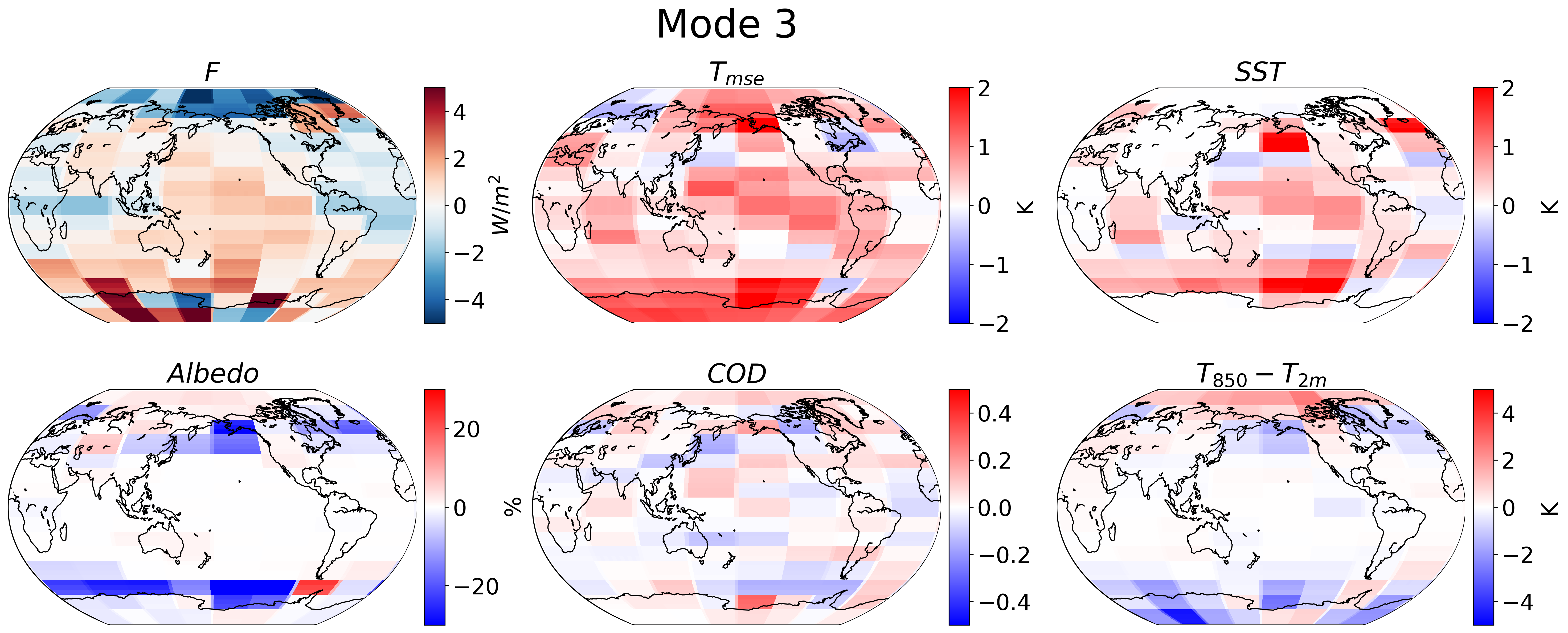}    
    \caption{As Fig. 4, but for the forcing-response pair of SVD mode-3.}
    \label{fig:mode-3}
\end{figure}
\clearpage 
\appendixtitle{Local contributions to Radiative Feedbacks}

\begin{figure}[h]
    \centering    \includegraphics[width=0.35\textwidth]{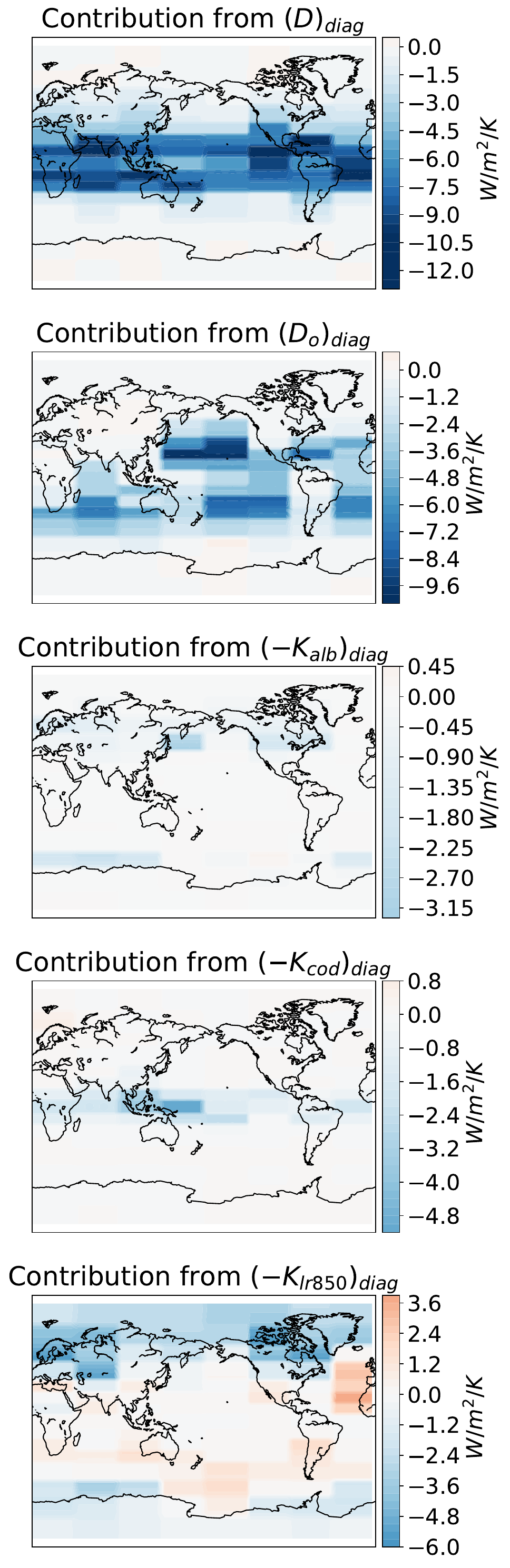}    
    \caption{Contribution to the forcing decomposition in Fig. \ref{fig:F_recon} from the diagonal components of $\mathcal{K}_m$.}
    \label{fig:mode-3}
\end{figure}
\clearpage

\bibliographystyle{ametsocV6}
\bibliography{references}

\begin{thebibliography}{74}
\providecommand{\natexlab}[1]{#1}
\providecommand{\url}[1]{\texttt{#1}}
\renewcommand{\UrlFont}{\rmfamily}
\providecommand{\urlprefix}{URL }
\expandafter\ifx\csname urlstyle\endcsname\relax
  \providecommand{\doi}[1]{https://doi.org/\discretionary{}{}{}#1}\else
  \providecommand{\doi}{https://doi.org/\discretionary{}{}{}\begingroup \urlstyle{rm}\Url}\fi
\providecommand{\eprint}[2][]{\url{#2}}

\bibitem[{Alexeev et~al.(2005)Alexeev, Langen,, and Bates}]{alexeev2005polar}
Alexeev, V., P.~Langen, and J.~Bates, 2005: Polar amplification of surface warming on an aquaplanet in “ghost forcing” experiments without sea ice feedbacks. \textit{Climate Dynamics}, \textbf{24}, 655--666.

\bibitem[{Andrews and Webb(2018)Andrews, and Webb}]{andrews2018dependence}
Andrews, T., and M.~J. Webb, 2018: The dependence of global cloud and lapse rate feedbacks on the spatial structure of tropical pacific warming. \textit{Journal of Climate}, \textbf{31~(2)}, 641--654.

\bibitem[{Armour et~al.(2013)Armour, Bitz,, and Roe}]{armour2013time}
Armour, K.~C., C.~M. Bitz, and G.~H. Roe, 2013: Time-varying climate sensitivity from regional feedbacks. \textit{Journal of Climate}, \textbf{26~(13)}, 4518--4534.

\bibitem[{Bates(2007)}]{bates2007some}
Bates, J., 2007: Some considerations of the concept of climate feedback. \textit{Quarterly Journal of the Royal Meteorological Society: A Journal of the Atmospheric sciences, Applied Meteorology and Physical Oceanography}, \textbf{133~(624)}, 545--560.

\bibitem[{Bates(2012)}]{bates2012climate}
Bates, J.~R., 2012: Climate stability and sensitivity in some simple conceptual models. \textit{Climate Dynamics}, \textbf{38}, 455--473.

\bibitem[{Beer and Eisenman(2022)Beer, and Eisenman}]{beer2022revisiting}
Beer, E., and I.~Eisenman, 2022: Revisiting the role of the water vapor and lapse rate feedbacks in the arctic amplification of climate change. \textit{Journal of Climate}, \textbf{35~(10)}, 2975--2988.

\bibitem[{Bintanja et~al.(2011)Bintanja, Graversen,, and Hazeleger}]{bintanja2011arctic}
Bintanja, R., R.~Graversen, and W.~Hazeleger, 2011: Arctic winter warming amplified by the thermal inversion and consequent low infrared cooling to space. \textit{Nature Geoscience}, \textbf{4~(11)}, 758--761.

\bibitem[{Black(1977)}]{black1977inventing}
Black, H.~S., 1977: Inventing the negative feedback amplifier: Six years of persistent search helped the author conceive the idea “in a flash” aboard the old {L}ackawanna ferry. \textit{IEEE spectrum}, \textbf{14~(12)}, 55--60.

\bibitem[{Bloch-Johnson et~al.(2021)Bloch-Johnson, Rugenstein, Stolpe, Rohrschneider, Zheng,, and Gregory}]{bloch2021climate}
Bloch-Johnson, J., M.~Rugenstein, M.~B. Stolpe, T.~Rohrschneider, Y.~Zheng, and J.~M. Gregory, 2021: Climate sensitivity increases under higher co2 levels due to feedback temperature dependence. \textit{Geophysical Research Letters}, \textbf{48~(4)}, e2020GL089\,074.

\bibitem[{Boer(2011)}]{boer2011ratio}
Boer, G., 2011: The ratio of land to ocean temperature change under global warming. \textit{Climate dynamics}, \textbf{37~(11)}, 2253--2270.

\bibitem[{Boer and Yu(2003{\natexlab{a}})Boer, and Yu}]{boer2003aclimate}
Boer, G., and B.~Yu, 2003{\natexlab{a}}: Climate sensitivity and climate state. \textit{Climate Dynamics}, \textbf{21~(2)}, 167--176.

\bibitem[{Boer and Yu(2003{\natexlab{b}})Boer, and Yu}]{boer2003bclimate}
Boer, G., and B.~Yu, 2003{\natexlab{b}}: Climate sensitivity and response. \textit{Climate Dynamics}, \textbf{20}, 415--429.

\bibitem[{Bony and Dufresne(2005)Bony, and Dufresne}]{bony2005marine}
Bony, S., and J.-L. Dufresne, 2005: Marine boundary layer clouds at the heart of tropical cloud feedback uncertainties in climate models. \textit{Geophysical Research Letters}, \textbf{32~(20)}.

\bibitem[{Caballero and Huber(2013)Caballero, and Huber}]{caballero2013state}
Caballero, R., and M.~Huber, 2013: State-dependent climate sensitivity in past warm climates and its implications for future climate projections. \textit{Proceedings of the National Academy of Sciences}, \textbf{110~(35)}, 14\,162--14\,167.

\bibitem[{Ceppi et~al.(2017)Ceppi, Brient, Zelinka,, and Hartmann}]{ceppi2017cloud}
Ceppi, P., F.~Brient, M.~D. Zelinka, and D.~L. Hartmann, 2017: Cloud feedback mechanisms and their representation in global climate models. \textit{Wiley Interdisciplinary Reviews: Climate Change}, \textbf{8~(4)}, e465.

\bibitem[{Colman and McAvaney(2009)Colman, and McAvaney}]{colman2009climate}
Colman, R., and B.~McAvaney, 2009: Climate feedbacks under a very broad range of forcing. \textit{Geophysical Research Letters}, \textbf{36~(1)}.

\bibitem[{Colman et~al.(1997)Colman, Power,, and McAvaney}]{colman1997non}
Colman, R., S.~Power, and B.~McAvaney, 1997: Non-linear climate feedback analysis in an atmospheric general circulation model. \textit{Climate dynamics}, \textbf{13}, 717--731.

\bibitem[{Colman and Soden(2021)Colman, and Soden}]{colman2021water}
Colman, R., and B.~J. Soden, 2021: Water vapor and lapse rate feedbacks in the climate system. \textit{Reviews of Modern Physics}, \textbf{93~(4)}, 045\,002.

\bibitem[{Crook et~al.(2011)Crook, Forster,, and Stuber}]{crook2011spatial}
Crook, J.~A., P.~M. Forster, and N.~Stuber, 2011: Spatial patterns of modeled climate feedback and contributions to temperature response and polar amplification. \textit{Journal of Climate}, \textbf{24~(14)}, 3575--3592.

\bibitem[{Dai et~al.(2019)Dai, Luo, Song,, and Liu}]{dai2019arctic}
Dai, A., D.~Luo, M.~Song, and J.~Liu, 2019: Arctic amplification is caused by sea-ice loss under increasing co2. \textit{Nature communications}, \textbf{10~(1)}, 121.

\bibitem[{Dong et~al.(2020)Dong, Armour, Zelinka, Proistosescu, Battisti, Zhou,, and Andrews}]{dong2020intermodel}
Dong, Y., K.~C. Armour, M.~D. Zelinka, C.~Proistosescu, D.~S. Battisti, C.~Zhou, and T.~Andrews, 2020: Intermodel spread in the pattern effect and its contribution to climate sensitivity in {CMIP5} and {CMIP6} models. \textit{Journal of Climate}, \textbf{33~(18)}, 7755--7775.

\bibitem[{Dong et~al.(2019)Dong, Proistosescu, Armour,, and Battisti}]{dong2019attributing}
Dong, Y., C.~Proistosescu, K.~C. Armour, and D.~S. Battisti, 2019: Attributing historical and future evolution of radiative feedbacks to regional warming patterns using a {G}reen’s function approach: The preeminence of the western {P}acific. \textit{Journal of Climate}, \textbf{32~(17)}, 5471--5491.

\bibitem[{Donohoe et~al.(2020)Donohoe, Armour, Roe, Battisti,, and Hahn}]{donohoe2020partitioning}
Donohoe, A., K.~C. Armour, G.~H. Roe, D.~S. Battisti, and L.~Hahn, 2020: The partitioning of meridional heat transport from the {L}ast {G}lacial maximum to {CO2} quadrupling in coupled climate models. \textit{Journal of Climate}, \textbf{33~(10)}, 4141--4165.

\bibitem[{England et~al.(2021)England, Eisenman, Lutsko,, and Wagner}]{england2021recent}
England, M.~R., I.~Eisenman, N.~J. Lutsko, and T.~J. Wagner, 2021: The recent emergence of arctic amplification. \textit{Geophysical Research Letters}, \textbf{48~(15)}, e2021GL094\,086.

\bibitem[{Faranda et~al.(2019)Faranda, Messori,, and Vannitsem}]{faranda2019attractor}
Faranda, D., G.~Messori, and S.~Vannitsem, 2019: Attractor dimension of time-averaged climate observables: insights from a low-order ocean-atmosphere model. \textit{Tellus A: Dynamic Meteorology and Oceanography}, \textbf{71~(1)}, 1554\,413.

\bibitem[{Feldl and Roe(2013)Feldl, and Roe}]{feldl2013four}
Feldl, N., and G.~Roe, 2013: Four perspectives on climate feedbacks. \textit{Geophysical Research Letters}, \textbf{40~(15)}, 4007--4011.

\bibitem[{Forster et~al.(2021)}]{forster2021earth}
Forster, P., and Coauthors, 2021: The earth’s energy budget, climate feedbacks, and climate sensitivity.

\bibitem[{Franzke et~al.(2005)Franzke, Majda,, and Vanden-Eijnden}]{franzke2005low}
Franzke, C., A.~J. Majda, and E.~Vanden-Eijnden, 2005: Low-order stochastic mode reduction for a realistic barotropic model climate. \textit{Journal of the atmospheric sciences}, \textbf{62~(6)}, 1722--1745.

\bibitem[{Goosse et~al.(2018)}]{goosse2018quantifying}
Goosse, H., and Coauthors, 2018: Quantifying climate feedbacks in polar regions. \textit{Nature communications}, \textbf{9~(1)}, 1919.

\bibitem[{Hahn et~al.(2021)Hahn, Armour, Zelinka, Bitz,, and Donohoe}]{hahn2021contributions}
Hahn, L.~C., K.~C. Armour, M.~D. Zelinka, C.~M. Bitz, and A.~Donohoe, 2021: Contributions to polar amplification in cmip5 and cmip6 models. \textit{Frontiers in Earth Science}, \textbf{9}, 710\,036.

\bibitem[{Held and Soden(2000)Held, and Soden}]{held2000water}
Held, I.~M., and B.~J. Soden, 2000: Water vapor feedback and global warming. \textit{Annual review of energy and the environment}, \textbf{25~(1)}, 441--475.

\bibitem[{Holland and Bitz(2003)Holland, and Bitz}]{holland2003polar}
Holland, M.~M., and C.~M. Bitz, 2003: Polar amplification of climate change in coupled models. \textit{Climate dynamics}, \textbf{21~(3)}, 221--232.

\bibitem[{Hu et~al.(2022)Hu, Xie,, and Kang}]{hu2022global}
Hu, S., S.-P. Xie, and S.~M. Kang, 2022: Global warming pattern formation: The role of ocean heat uptake. \textit{Journal of Climate}, \textbf{35~(6)}, 1885--1899.

\bibitem[{Huang and Huang(2021)Huang, and Huang}]{huang2021nonlinear}
Huang, H., and Y.~Huang, 2021: Nonlinear coupling between longwave radiative climate feedbacks. \textit{Journal of Geophysical Research: Atmospheres}, \textbf{126~(8)}, e2020JD033\,995.

\bibitem[{Huang et~al.(2021)Huang, Huang,, and Hu}]{huang2021quantifying}
Huang, H., Y.~Huang, and Y.~Hu, 2021: Quantifying the energetic feedbacks in {ENSO}. \textit{Climate Dynamics}, \textbf{56}, 139--153.

\bibitem[{Huang et~al.(2017)Huang, Xia,, and Tan}]{huang2017pattern}
Huang, Y., Y.~Xia, and X.~Tan, 2017: On the pattern of co2 radiative forcing and poleward energy transport. \textit{Journal of Geophysical Research: Atmospheres}, \textbf{122~(20)}, 10--578.

\bibitem[{Hummel et~al.(2023)Hummel, Ashwin,, and Kuehn}]{hummel2023reduction}
Hummel, F., P.~Ashwin, and C.~Kuehn, 2023: Reduction methods in climate dynamics—a brief review. \textit{Physica D: Nonlinear Phenomena}, \textbf{448}, 133\,678.

\bibitem[{James et~al.(2013)James, Witten, Hastie, Tibshirani et~al.}]{james2013introduction}
James, G., D.~Witten, T.~Hastie, R.~Tibshirani, and Coauthors, 2013: \textit{An introduction to statistical learning}, Vol. 112. Springer.

\bibitem[{Jonko et~al.(2012)Jonko, Shell, Sanderson,, and Danabasoglu}]{jonko2012climate}
Jonko, A.~K., K.~M. Shell, B.~M. Sanderson, and G.~Danabasoglu, 2012: Climate feedbacks in ccsm3 under changing co2 forcing. part i: Adapting the linear radiative kernel technique to feedback calculations for a broad range of forcings. \textit{Journal of Climate}, \textbf{25~(15)}, 5260--5272.

\bibitem[{Jonko et~al.(2013)Jonko, Shell, Sanderson,, and Danabasoglu}]{jonko2013climate}
Jonko, A.~K., K.~M. Shell, B.~M. Sanderson, and G.~Danabasoglu, 2013: Climate feedbacks in ccsm3 under changing co 2 forcing. part ii: Variation of climate feedbacks and sensitivity with forcing. \textit{Journal of Climate}, \textbf{26~(9)}, 2784--2795.

\bibitem[{Kay et~al.(2012)Kay, Holland, Bitz, Blanchard-Wrigglesworth, Gettelman, Conley,, and Bailey}]{kay2012influence}
Kay, J.~E., M.~M. Holland, C.~M. Bitz, E.~Blanchard-Wrigglesworth, A.~Gettelman, A.~Conley, and D.~Bailey, 2012: The influence of local feedbacks and northward heat transport on the equilibrium arctic climate response to increased greenhouse gas forcing. \textit{Journal of Climate}, \textbf{25~(16)}, 5433--5450.

\bibitem[{Klein and Hartmann(1993)Klein, and Hartmann}]{klein1993seasonal}
Klein, S.~A., and D.~L. Hartmann, 1993: The seasonal cycle of low stratiform clouds. \textit{Journal of Climate}, \textbf{6~(8)}, 1587--1606.

\bibitem[{Knutti and Sedl{\'a}{\v{c}}ek(2013)Knutti, and Sedl{\'a}{\v{c}}ek}]{knutti2013robustness}
Knutti, R., and J.~Sedl{\'a}{\v{c}}ek, 2013: Robustness and uncertainties in the new cmip5 climate model projections. \textit{Nature climate change}, \textbf{3~(4)}, 369--373.

\bibitem[{Lin et~al.(2021)Lin, Hwang, Lu, Liu,, and Rose}]{lin2021dominant}
Lin, Y.-J., Y.-T. Hwang, J.~Lu, F.~Liu, and B.~E. Rose, 2021: The dominant contribution of southern ocean heat uptake to time-evolving radiative feedback in cesm. \textit{Geophysical Research Letters}, \textbf{48~(9)}, e2021GL093\,302.

\bibitem[{Liu et~al.(2018)Liu, Lu, Garuba, Huang, Leung, Harrop,, and Luo}]{liu2018sensitivity}
Liu, F., J.~Lu, O.~A. Garuba, Y.~Huang, L.~R. Leung, B.~E. Harrop, and Y.~Luo, 2018: Sensitivity of surface temperature to oceanic forcing via q-flux green’s function experiments. {P}art {II}: {F}eedback decomposition and polar amplification. \textit{Journal of Climate}, \textbf{31~(17)}, 6745--6761.

\bibitem[{Liu et~al.(2020)Liu, Lu, Huang, Leung, Harrop,, and Luo}]{liu2020sensitivity}
Liu, F., J.~Lu, Y.~Huang, L.~R. Leung, B.~E. Harrop, and Y.~Luo, 2020: Sensitivity of surface temperature to oceanic forcing via q-flux green’s function experiments. part iii: asymmetric response to warming and cooling. \textit{Journal of Climate}, \textbf{33~(4)}, 1283--1297.

\bibitem[{Liu et~al.(2022)Liu, Lu,, and Leung}]{liu2022neutral}
Liu, F., J.~Lu, and L.~R. Leung, 2022: Neutral mode dominates the forced global and regional surface temperature response in the past and future. \textit{Geophysical Research Letters}, \textbf{49~(15)}, e2022GL098\,788.

\bibitem[{Lu and Cai(2009)Lu, and Cai}]{lu2009new}
Lu, J., and M.~Cai, 2009: A new framework for isolating individual feedback processes in coupled general circulation climate models. part i: Formulation. \textit{Climate dynamics}, \textbf{32}, 873--885.

\bibitem[{Lu et~al.(2020)Lu, Liu, Leung,, and Lei}]{lu2020neutral}
Lu, J., F.~Liu, L.~R. Leung, and H.~Lei, 2020: Neutral modes of surface temperature and the optimal ocean thermal forcing for global cooling. \textit{npj Climate and Atmospheric Science}, \textbf{3~(1)}, 9.

\bibitem[{Lucarini et~al.(2016)}]{lucarini2016extremes}
Lucarini, V., and Coauthors, 2016: \textit{Extremes and recurrence in dynamical systems}. John Wiley \& Sons.

\bibitem[{Majda et~al.(2008)Majda, Franzke,, and Khouider}]{majda2008applied}
Majda, A.~J., C.~Franzke, and B.~Khouider, 2008: An applied mathematics perspective on stochastic modelling for climate. \textit{Philosophical Transactions of the Royal Society A: Mathematical, Physical and Engineering Sciences}, \textbf{366~(1875)}, 2427--2453.

\bibitem[{Merlis and Henry(2018)Merlis, and Henry}]{merlis2018simple}
Merlis, T.~M., and M.~Henry, 2018: Simple estimates of polar amplification in moist diffusive energy balance models. \textit{Journal of Climate}, \textbf{31~(15)}, 5811--5824.

\bibitem[{Neelin et~al.(2006)Neelin, M{\"u}nnich, Su, Meyerson,, and Holloway}]{neelin2006tropical}
Neelin, J.~D., M.~M{\"u}nnich, H.~Su, J.~E. Meyerson, and C.~E. Holloway, 2006: Tropical drying trends in global warming models and observations. \textit{Proceedings of the National Academy of Sciences}, \textbf{103~(16)}, 6110--6115.

\bibitem[{Pendergrass et~al.(2018)Pendergrass, Conley,, and Vitt}]{pendergrass2018surface}
Pendergrass, A.~G., A.~Conley, and F.~M. Vitt, 2018: Surface and top-of-atmosphere radiative feedback kernels for cesm-cam5. \textit{Earth System Science Data}, \textbf{10~(1)}, 317--324.

\bibitem[{Pithan and Mauritsen(2014)Pithan, and Mauritsen}]{pithan2014arctic}
Pithan, F., and T.~Mauritsen, 2014: Arctic amplification dominated by temperature feedbacks in contemporary climate models. \textit{Nature geoscience}, \textbf{7~(3)}, 181--184.

\bibitem[{Previdi et~al.(2021)Previdi, Smith,, and Polvani}]{previdi2021arctic}
Previdi, M., K.~L. Smith, and L.~M. Polvani, 2021: Arctic amplification of climate change: a review of underlying mechanisms. \textit{Environmental Research Letters}, \textbf{16~(9)}, 093\,003.

\bibitem[{Proistosescu and Huybers(2017)Proistosescu, and Huybers}]{proistosescu2017slow}
Proistosescu, C., and P.~J. Huybers, 2017: Slow climate mode reconciles historical and model-based estimates of climate sensitivity. \textit{Science Advances}, \textbf{3~(7)}, e1602\,821.

\bibitem[{Roe(2009)}]{roe2009feedbacks}
Roe, G., 2009: Feedbacks, timescales, and seeing red. \textit{Annual Review of Earth and Planetary Sciences}, \textbf{37}, 93--115.

\bibitem[{Roe et~al.(2015)Roe, Feldl, Armour, Hwang,, and Frierson}]{roe2015remote}
Roe, G.~H., N.~Feldl, K.~C. Armour, Y.-T. Hwang, and D.~M. Frierson, 2015: The remote impacts of climate feedbacks on regional climate predictability. \textit{Nature Geoscience}, \textbf{8~(2)}, 135--139.

\bibitem[{Rose and Rayborn(2016)Rose, and Rayborn}]{rose2016effects}
Rose, B.~E., and L.~Rayborn, 2016: The effects of ocean heat uptake on transient climate sensitivity. \textit{Current climate change reports}, \textbf{2}, 190--201.

\bibitem[{Rugenstein et~al.(2020)}]{rugenstein2020equilibrium}
Rugenstein, M., and Coauthors, 2020: Equilibrium climate sensitivity estimated by equilibrating climate models. \textit{Geophysical Research Letters}, \textbf{47~(4)}, e2019GL083\,898.

\bibitem[{Screen and Simmonds(2010)Screen, and Simmonds}]{screen2010central}
Screen, J.~A., and I.~Simmonds, 2010: The central role of diminishing sea ice in recent arctic temperature amplification. \textit{Nature}, \textbf{464~(7293)}, 1334--1337.

\bibitem[{Shell et~al.(2008)Shell, Kiehl,, and Shields}]{shell2008using}
Shell, K.~M., J.~T. Kiehl, and C.~A. Shields, 2008: Using the radiative kernel technique to calculate climate feedbacks in ncar’s community atmospheric model. \textit{Journal of Climate}, \textbf{21~(10)}, 2269--2282.

\bibitem[{Shepherd(2014)}]{shepherd2014atmospheric}
Shepherd, T.~G., 2014: Atmospheric circulation as a source of uncertainty in climate change projections. \textit{Nature Geoscience}, \textbf{7~(10)}, 703--708.

\bibitem[{Smith et~al.(2020)Smith, Kramer,, and Sima}]{smith2020hadgem3}
Smith, C.~J., R.~J. Kramer, and A.~Sima, 2020: The hadgem3-ga7. 1 radiative kernel: the importance of a well-resolved stratosphere. \textit{Earth System Science Data}, \textbf{12~(3)}, 2157--2168.

\bibitem[{Soden and Held(2006)Soden, and Held}]{soden2006assessment}
Soden, B.~J., and I.~M. Held, 2006: An assessment of climate feedbacks in coupled ocean--atmosphere models. \textit{Journal of climate}, \textbf{19~(14)}, 3354--3360.

\bibitem[{Soden et~al.(2008)Soden, Held, Colman, Shell, Kiehl,, and Shields}]{soden2008quantifying}
Soden, B.~J., I.~M. Held, R.~Colman, K.~M. Shell, J.~T. Kiehl, and C.~A. Shields, 2008: Quantifying climate feedbacks using radiative kernels. \textit{Journal of Climate}, \textbf{21~(14)}, 3504--3520.

\bibitem[{Taylor et~al.(2022)}]{taylor2022process}
Taylor, P.~C., and Coauthors, 2022: Process drivers, inter-model spread, and the path forward: A review of amplified arctic warming. \textit{Frontiers in Earth Science}, \textbf{9}, 758\,361.

\bibitem[{Winton(2006)}]{winton2006amplified}
Winton, M., 2006: Amplified arctic climate change: What does surface albedo feedback have to do with it? \textit{Geophysical Research Letters}, \textbf{33~(3)}.

\bibitem[{Xie et~al.(2015)}]{xie2015towards}
Xie, S.-P., and Coauthors, 2015: Towards predictive understanding of regional climate change. \textit{Nature Climate Change}, \textbf{5~(10)}, 921--930.

\bibitem[{Zelinka and Hartmann(2012)Zelinka, and Hartmann}]{zelinka2012climate}
Zelinka, M.~D., and D.~L. Hartmann, 2012: Climate feedbacks and their implications for poleward energy flux changes in a warming climate. \textit{Journal of Climate}, \textbf{25~(2)}, 608--624.

\bibitem[{Zhang et~al.(1994)Zhang, Hack, Kiehl,, and Cess}]{zhang1994diagnostic}
Zhang, M., J.~Hack, J.~Kiehl, and R.~Cess, 1994: Diagnostic study of climate feedback processes in atmospheric general circulation models. \textit{Journal of Geophysical Research: Atmospheres}, \textbf{99~(D3)}, 5525--5537.

\bibitem[{Zhang et~al.(2018)Zhang, Wang, Fu, Pendergrass, Wang, Yang, Ma,, and Rasch}]{zhang2018local}
Zhang, R., H.~Wang, Q.~Fu, A.~G. Pendergrass, M.~Wang, Y.~Yang, P.-L. Ma, and P.~J. Rasch, 2018: Local radiative feedbacks over the arctic based on observed short-term climate variations. \textit{Geophysical Research Letters}, \textbf{45~(11)}, 5761--5770.

\bibitem[{Zhou et~al.(2016)Zhou, Zelinka,, and Klein}]{zhou2016impact}
Zhou, C., M.~D. Zelinka, and S.~A. Klein, 2016: Impact of decadal cloud variations on the earth’s energy budget. \textit{Nature Geoscience}, \textbf{9~(12)}, 871--874.

\end{thebibliography}

\end{document}